\documentclass[12pt]{article}
\usepackage{amsmath}
\usepackage{graphicx,psfrag,epsf}
\usepackage{enumerate}
\usepackage{url} 

\usepackage{amsthm}
\usepackage{amsmath}
\usepackage{amssymb}
\usepackage{mathtools}

\usepackage{xcolor}

\usepackage{rotating}
\usepackage{pdflscape}
\usepackage{graphicx}

\usepackage{booktabs}
\usepackage{siunitx}

\usepackage{comment}

\usepackage{bibentry}
\usepackage{appendix}

\usepackage{enumitem}

\usepackage{times}
\usepackage{bm}
\usepackage{natbib}
\usepackage{setspace}
\def\bSig\mathbf{\Sigma}

\makeatletter
\newcommand*\bigcdot{\mathpalette\bigcdot@{.5}}
\newcommand*\bigcdot@[2]{\mathbin{\vcenter{\hbox{\scalebox{#2}{$\m@th#1\bullet$}}}}}
\makeatother

\def\Z{{\boldsymbol{Z}}}

\def\U{{\boldsymbol{U}}}
\def\W{{\boldsymbol{W}}}

\def\l{\left}
\def\r{\right}
\def\R{\mathbb{R}}
\def\mc{\mathcal}

\begin{document}

\def\spacingset#1{\renewcommand{\baselinestretch}%
{#1}\small\normalsize} \spacingset{1}


\bigskip
\bigskip
\bigskip
\begin{center}
{\LARGE Risk-Adjusted Incidence Modeling on Hierarchical Survival Data with Recurrent Events}
\end{center}
\bigskip
\begin{center}{X.Jiang${}^1$, W.Stoudemire${}^2$, M.S.Muhlebach${}^2$, M.R.Kosorok${}^1$}\end{center}

\begin{center}
\emph{${}^1$Department of Biostatistics,  University of North Carolina, Chapel Hill, NC\\${}^2$Department of Pediatrics,	University of North Carolina, Chapel Hill, NC}
\end{center}
\medskip

\bigskip
\begin{abstract}
There is a constant need for many healthcare programs to timely address problems with infection prevention and control (IP\&C). For example, pathogens can be transmitted among patients with cystic fibrosis (CF) in both the inpatient and outpatient settings within the healthcare system even with the existing recommended IP\&C practices, and these pathogens are often associated with negative clinical outcomes. Because of limited and delayed data sharing, CF programs need a reliable method to track infection rates. There are three complex structures in CF registry data: recurrent infections, missing data, and multilevel correlation due to repeated measures within a patient and patient-to-patient transmissions. A step-by-step analysis pipeline was proposed to develop and validate a risk-adjusted model to help healthcare programs monitor the number of recurrent events while taking into account missing data and the hierarchies of repeated measures in right-censored data. We extended the mixed-effect Andersen-Gill model (the frailty model), adjusted for important risk factors, and provided confidence intervals for the predicted number of events where the variability of the prediction was estimated from three identified sources. The coverage of the estimated confidence intervals was used to evaluate model performance. Simulation results indicated that the coverage of our method was close to the desired confidence level. To demonstrate its clinical practicality, our pipeline was applied to monitor the infection incidence rate of two key CF pathogens using a U.S. registry. Results showed that years closer to the time of interest were better at predicting future incidence rates in the CF example.
\end{abstract}

\noindent%
\vfill

\newpage
\spacingset{1.45} 

\section{Introduction}
\label{sec:survival-intro}

Right-censored data such as patient encounters in healthcare settings often inevitably have missing data and a multilevel structure where the assumption of independent observations is violated. Moreover, there is a need for many health institutions to monitor infection events which could occur repeatedly for patients. Motivated by the current situation of infection tracking in cystic fibrosis, we expanded on existing statistical approaches in mixed effect survival models and resampling methods to provide program-specific predictions and estimated confidence intervals (CIs) with desired coverage. The proposed prediction tool adjusts for various sources of variability in the complex data and provides healthcare programs with a practical way to detect excessive or fewer-than-expected events.

In healthcare infection control, the primary summary statistic to track healthcare-associated infections is the standardized infection ratio  (SIR).  
This ratio is defined as the observed number of infections divided by the expected number of infections, the latter of which is a summation of the number of patients weighted by the national standard stratum-specific rates  \citep{centers2018nhsn, gustafson2006three}. By the Centers for Disease Control and Prevention (CDC) guide \citep{sirguide}, 
the number of predicted infections is calculated from a logistic or negative binomial regression adjusting for known risk factors associated with infections, such as patient characteristics and geography. A ratio less than 1 means the incident infections are fewer than expected and a ratio more than 1 means the incident infections are worse than expected. SIR has several limitations that is not ideal for different types of data and applications \citep{gustafson2006three}. Although it has demonstrated that risk adjustment methods are promising, SIR compares infections at a healthcare program with the national benchmark at a single time point. Since the ratio is based on cross-sectional nationwide expectations, it does not adjust for right censoring and is not ideal for comparisons between hospitals or across time \citep{delgado2005caution}. Infection is one of the examples of lifetime data where time and censoring play an important role in risk adjustment that needs appropriate methodology. Inspired by this clinical background, we aimed to improve the way to predict and monitor recurrent events in complex health data that would allow us to compare between healthcare programs while taking into account repeated measures in the presence of missing data.

Our method design was based on the CF Foundation Patient Registry (CFFPR) from which three complex data structures had been identified: recurrent infections, multilevel hierarchy, and missing data. There is a diverse literature of survival models that tackle recurrent survival events with multilevel clustering  or missing data. Some methods \citep{yau2001multilevel, mcgilchrist1993reml} use random effects to model the multilevel clustering by incorporating a Cox model with a generalized linear mixed effect model and obtaining REML estimators for variance. Bayesian joint modeling can learn recurrent events among highly heterogeneous patients creating robust estimators without parametric constraints \citep{r2003bayesian}. Such Bayesian approach has been extended to multilevel clustering data to account for clinical site correlation or lesion correlation \citep{luo2014bayesian, brilleman2019joint} by novel patient-specific association structure. Multiple imputation has been used for missing covariates in time-to-event or multilevel data \citep{van1999multiple, van2011multiple} in general. For example, multiple imputation can be used to impute interval-censored event times in hierarchical data before fitting a frailty model \cite{lam2010multiple}. These methods focus on recurrent events, hierarchical repeated measures, missing data, or a combination of the two but do not necessarily address all three problems at the same time.

This paper has two main goals: i) To develop a risk-adjusted pipeline for predicting program-level events for right-censored repeated event data while incorporating complex situations such as multilevel hierarchical clustering and missing data; ii) To discover the validity and practicality of this model on the incidence of bacterial infection using U.S. nation-wide registry data from the Cystic Fibrosis (CF) Foundation. 

The rest of the paper is organized as follows: We generalize the problem and layout modeling methods as a pipeline in Section \ref{sec:survival_methods}: we review existing survival models in Section \ref{sec:survival_survmodel}, set up our problem of interest in Section \ref{sec:survival_setup}, and provide explanation of the parameter estimation and its variability estimation in Sections \ref{sec:survival_estandvar} \& \ref{sec:survival_theojust}. Appendix D contains a glossary of all notations and symbols introduced in Section \ref{sec:survival_methods}. Numerical experiments are explored in Section \ref{sec:survival_sims} to investigate the model performance. In Section \ref{sec:survival_clinicalimp}, we demonstrate the implementation of the proposed methods with CF data, illustrating each step of the analysis pipeline. Strengths, limitations, and future research are discussed in Section \ref{sec:survival_discuss}. 

\section{Methods}\label{sec:survival_methods}

\subsection{The Mixed Effect Andersen-Gill Model}\label{sec:survival_survmodel}

When there are multilevel hierarchies in recurrent event data, the standard Cox proportional hazards model is no longer suitable because it is designed to assess time to the first event. Alternative models need to be considered to accommodate for these special features. The Andersen-Gill (AG) model \citep{andersen1982cox} is a popular choice that extends the common Cox model to repeated time-to-event data. However, AG assumes that the recurrent event times are independent conditioning on time-varying covariates \citep{amorim2015modelling} and the baseline intensity is the same across all recurrent events \citep{yang2017statistical}. These assumptions do not necessarily hold in the hierarchical data setting we are interested in. Other models that also address multiple failure times include Prentice-William-Peterson (PWP) and Wei-Lin-Weissfeld (WLW), which are well-developed and robust but do not explore the relationships between failures \citep{wei1997overview}. There are two leading ways to model both recurrent events and hierarchical structure: 1) A marginal approach: a generalized estimating equation (GEE) to estimate the parameters of the marginal cumulative incidence function where the correlated observations are taken into account by the covariance matrix with a sandwich estimator \citep{logan2011marginal}. This approach focuses on the marginal covariate effect on failure risks by adjusting the variance but not necessarily the coefficients; 
2) A conditional approach: a mixed effect survival model where random effects can be incorporated as a frailty model and the baseline hazard varies by the group variable(s), resulting in a multiplicative effect on the hazard. This approach adjusts for both coefficients and the covariance. Conditioning on the random effects, it is assumed that the intensity function of each subject follows the AG model \citep{wei1997overview}; thus this approach can be deemed as a mixed effect AG model. 

The second approach was used in this study because we were more interested in the effect on the individual-level instead of population-level (individual can be broad here; it could be a subject or a healthcare program) and we believe the inclusion of random effects can improve the model fit and CIs. 
The random effects describe the term `frailty', which is the excessive risk for distinct grouping variables \citep{therneau2003penalized}. In general, the hazard of a frailty model for subject $i$ in group $j$ is
$
\lambda_{ij}(t) =   \lambda_0(t) e^{\U_i \beta + \W_{ij} b} = \lambda_0(t) e^{\U_{ij} \beta}  \omega_{ij},$
where $\lambda_0$ is baseline hazard, $\U_i$ denotes covariates for the $i$th subject, $\W_{ij} = 1$ if subject $i$ belongs to the $j$th group, $\beta, b$ are the fixed and random effects, and $\omega_{ij} = \exp(\W_{ij} b)$ is the unmeasured frailty  \citep{pickles1995comparison}. The frailty term accounts for variation in the risk that is not captured solely by the covariates, and the frailty model assumes that the time increments are uncorrelated once we adjust for both the covariates and random effects \citep{pickles1995comparison, amorim2015modelling}. Our setup meets this assumption. Intuitively speaking, we assume that there is unobserved information (i.e., the random effects) that explains the heterogeneity in the data which cannot be explained only by the observed covariates. In addition, the linear mixed effect approach follows the partial likelihood approach \cite{cox1975partial} where the key advantage is that the maximum partial likelihood estimation of the covariates requires the baseline hazard function to be specified and is unbiased and asymptotically normally distributed under mild conditions \citep{yau2001multilevel}. 

\subsection{The Setup and Overview}\label{sec:survival_setup}

Assume we have two hierarchical levels in the right censored data. There are $N$ unique level-1 groups (the highest hierarchical level) denoted by $j = 1, \ldots, N$. Within the $j$th level-1 group, there are $c_j$ unique level-2 groups where $i = 1,\ldots, c_j$. Observations in the $i$th level-2 group and $j$th level-1 group are denoted by $k = 1, \ldots, m_{ij}$. 
For example, an event in CF could be a bacterial infection incidence, which is recurrent given a reasonable washout period. Each CF program is a level-1 group $j$, the CF patient is a level-2 group $i$, and the number of encounters for a CF patient who goes to a CF program is represented by $k$. 
Potential risk factors such as program and patient characteristics for the $j$th CF program, $i$th CF patient, and $k$th encounter are denoted by $\Z_{jik}(t) \in \R^p$ at time $t$ where $p$ is the total number of covariates of interest. $\Z$ represents both time-varying and time-invariant covariates but  time-invariant covariates have constant values across $t$. The at-risk time interval is denoted by $T_{jik}$. The true intensity process of the counting process is 
\begin{equation}\label{eqn:true_intensity}
R_{0j} = \sum_{i=1}^{c_j} \sum_{k=1}^{m_{ij}} \int_0^{T_{jik}} e^{\Z_{jik}(s)\beta_0 + b_{ij}} d\Lambda_0(s),
\end{equation}
for each program $j$ \citep{lawless1987regression}. The random effects $b_{ij}$'s are i.i.d. zero-mean Gaussian random variables. Because level-1 and level-2 groups can be completely nested or crossed, we used the general notation $b_{ij}$ here to capture random effects from both groups (denoted by $i$ and $j$). The covariate effect $\beta_0$ was estimated by the partial likelihood on the full $n$ dataset, i.e. all
events for all patients at all programs, where $n = \sum_{j=1}^{N} \sum_{i=1}^{c_j} m_{ij} $. Given the covariate effect estimator $\widehat{\beta}_n$, the hazard was estimated with an extension of the Breslow estimator \citep{lin2007breslow}. The variability of the estimators was estimated with the block jackknife method \citep{ma2005penalized}. The predicted number of events for each program comes with a two-sided CI, 
calculated based on the estimated intensity process and the Poisson process that describes the events. 
We validated the model estimated in the training period with a separate dataset, and compared the estimated, risk-adjusted CI with the true, observed number of events to evaluate the accuracy of the prediction model. Figure 1 
describes the whole pipeline of our data analysis method. 

\subsection{Estimation of Parameters and Their Variability}\label{sec:survival_estandvar}

Missing data have become a universal issue in data analysis due to non-responses and data collection mistakes or simply because the information wanted is not available. Since it could cause potential problems if left untended, many researchers resort to imputation. We chose multiple imputation (MI) because we had a not negligible proportion of missing data, the missing completely at random assumption was not plausible in the CF example, and MI accounted for the uncertainty in the data. Recall that $\beta$ is a $p$-dimensional vector with $p$ being the number of covariates in the data. Assume there are $M$ copies of MI and let $\hat\beta_{nl}$ be the estimated coefficient from the frailty model for the $l$th MI dataset ($l=1, \ldots, M$) trained on a data size of $n$. We applied the Rubin's rule \citep{rubin2004multiple} and pooled the multiple imputed $\hat\beta_{nl}$ together to determine important risk factors. The pooled beta estimate is the average over $\hat{\beta}_{nl}$'s across $M$ MI copies $$\bar{\hat{\beta}}_M = \frac{1}{M} \sum_{l=1}^M \hat\beta_{nl},$$ and the variance of this pooled estimate is $$Var(\bar{\hat{\beta}}_M) = \frac{1}{M} \sum_{l=1}^M \text{Cov}(\hat\beta_{nl}) + \l(1+\frac{1}{M}\r) \frac{\sum_{l=1}^M (\hat\beta_{nl} - \bar{\hat{\beta}}_M)^T (\hat\beta_{nl} - \bar{\hat{\beta}}_M)}{M-1},$$ where the first term is the average of the variance-covariance of $\hat\beta_{nl}$ estimate in the fitted mixed effect AG model (within MI) and the second term is the variance across the MI estimates $\hat\beta_{nl}$'s (across MI). The test statistic for the pooled estimate is then defined as $T_{pooled} = \frac{\bar{\hat{\beta}}_M}{\l(Var(\bar{\hat{\beta}}_M)\r)^{1/2}} $ whose null distribution asymptotically follows a two-sided student t-distribution. 

Besides $\hat\beta_{nl}$, another estimator we needed is the estimated hazard function. We extended the Breslow estimator \citep{lin2007breslow} and defined the hazard function estimator as the number of empirical events across both levels of hierarchy at a certain time point adjusted by the at-risk population whose at risk interval contains the current time point. Mathematically, the baseline hazard function estimator is
\begin{equation}\label{eqn:dlambda}
d\hat\Lambda_{nl}(s) = \frac{\sum_{j=1}^N \sum_{i=1}^{c_j} \sum_{k=1}^{m_{ij}} d N_{jik}(s)}{\sum_{j=1}^N \sum_{i=1}^{c_j} \sum_{k=1}^{m_{ij}} 1\{L_{jik} < s \leq U_{jik}\} e^{\Z_{jikl}(s)\hat\beta_{nl} }},
\end{equation}
where $j,i,k$ were introduced in Section \ref{sec:survival_setup}, $dN_{jik}(s)$ is the indicator of an observed recurrent survival event at time $s$ (the instantaneous change in the counting process), $\Z_{jikl}(s)$ represents covariates at time $s$ for the $l$th MI copy, and $L_{jik}, U_{jik}$ are the lower and upper bounds of the at-risk interval for the $j$th level-1 group, $i$th level-2 group, and $k$th encounter. 
Because we cared more about the relative length, the at-risk intervals are shifted to relative at-risk days rather than the actual dates so the first interval for each subject starts with zero. 

When there are multiple hierarchies in the right-censored data, which level of the hierarchy is of interest? The purpose of this study is to monitor and predict the number of events at the program level (e.g., the highest level). This is a common task especially for the field of infection prevention and control (IP\&C) where we are interested in how each healthcare program is doing as a whole rather than how well each patient is.
The observed number of events at the $j$th level-1 group (e.g., CF program $j$) is summed over all its level-2 groups (e.g., CF patients $i$) and encounters ($k$):
\begin{equation} \label{eqn:trueN}
N_j = \sum_{i=1}^{c_j} \sum_{k=1}^{m_{ij}} N_{jik} \ .
\end{equation} 
The expected number of risk-adjusted events for the $j$th level-1 group and $l$th MI is 
\begin{equation}\label{eqn:N.hat.jM} \hat{N}_{jl}
= \sum_{i=1}^{c_j} \sum_{k=1}^{m_{ij}} \int_s 1\{L_{jik} < s \leq U_{jik}\} e^{Z_{jikl}(s) \hat\beta_{nl}} d\hat\Lambda_{nl}(s).
\end{equation}
We obtained the expected number of risk-adjusted events by averaging over all MIs:
\begin{equation} \label{eqn:N.hat.j}
\hat{N}_{j \bigcdot} = \frac{1}{M}\sum_{l=1}^M \hat{N}_{jl}.
\end{equation}
The variation of this estimator mainly comes from the two plug-in estimators, $\hat\beta_{nl}$ and $d\hat\Lambda_{nl}$. We identified three components that contribute to the variability of the estimated number of events, $\widehat{\widetilde{V}}_{j}$:
\begin{enumerate}
	\item[1.] Within-group variance: $\hat{N}_{j\bigcdot} \ $. The variance of estimated events for all level-2 groups ($i$'s) within each level-1 group ($j$). The ``group'' in ``within-group'' refers to the level-1 group;
	\item[2.] Across-group variance: $\hat{V}_{j\bigcdot} = \frac{1}{M} \sum_{l=1}^M  \hat{V}\l(\hat{N}_{jl}\r)$. The variance of estimated events across level-1 groups ($i$'s);
	\item[3.] Multiple imputation variance: $\hat{s}_{j\bigcdot}^2 = \frac{1}{M} \sum_{l=1}^M \l( \hat{N}_{jl} - \hat{N}_{j\bigcdot}\r)^2$. The variance of estimated events due to multiple imputed copies. 
\end{enumerate}
Thus, 
\begin{equation}\label{eqn:V.hat.tilde.j}
\widehat{\widetilde{V}}_{j} = \hat{N}_{j\bigcdot} + \hat{V}_{j\bigcdot} + \hat{s}_{j\bigcdot}^2 \quad. 
\end{equation}
The first and third components are readily available since we know $\hat{N}_{jl}$ and $\hat{N}_{j\bigcdot}$. The second component, the across-group variance, is estimated using the block jackknife. Compared with bootstrap, block jackknife is a computationally simpler method which has fewer assumptions and is a a valid, robust variance estimator even for mis-specified models \citep{ma2005penalized}.
Let a fixed integer $m$ be the number of blocks and $q_{m,N}$ be the number of elements in each block, which is defined as the largest integer such that $m \cdot q_{m,N} \leq N,$ where $N$ is the unique number of values of the level-1 grouping variable. For example, if there are $N=271$ level-1 groups and $m = 10$ blocks, each block will contain $q_{m,N} = 27$ elements. 
Next, the $m \cdot q_{m,N}$ observations are randomly sampled from the original data $D$, denoted as $D^*$, to prepare for splitting the data into blocks of equal sizes. The notation $*$ is used to distinguish the block jackknife data from the original data. For each block $b = 1, \ldots, m$, ${{\hat{\theta^*}}_j}^{(-b)} = \hat{N}_j \l({\hat{\beta^*}}^{(-b)}, {d\hat{\Lambda^*}}^{(-b)} \r)$ is obtained based on ${D^*}^{(-b)}$, which is the $m \cdot q_{m,N}$ randomly sampled level-1 groups after omitting the $b$th block (hence $b-1$ blocks remaining). We combined the estimators and their estimated variances across all $m$ blocks: $$\bar{\theta}_j^* = \frac{1}{m} \sum_{b=1}^m {\hat{\theta^*}}_j^{(-b)} \quad \text{    and    } \quad 
S_j^* = (m-1) q_{m,N} \sum_{b=1}^m (\hat{\theta}_j^{*^{(-b)}} - \bar{\theta}_j^*)(\hat{\theta}_j^{*^{(-b)}} - \bar{\theta}_j^*)^T.$$
Note that the MI is omitted here for simplicity, this block jackknife algorithm should be repeated for each $l = 1, \ldots, M$ multiple imputed copies with $l$ added to the subscript of $\bar{\theta}_j^*$ and $S^*_j$. The second component of the variance of $\widehat{\widetilde{V}}_{j}$ is then $$\hat{V}_{j\bigcdot} = \frac{1}{M} \sum_{l=1}^M \hat{V}_{jl} =  \frac{1}{M} \sum_{l=1}^M S^*_{jl}.$$ More justifications of the three components 
and the block jackknife method are located in Section \ref{sec:survival_theojust}. 

Now that we have all components, we need to construct a hypothesis testing how precise our event estimator is compared to the true number of events, adjusted for the variability. The test statistic of a Z-test is
\begin{equation} \label{eqn:test_stat}
\hat{T}_j = \frac{N_j - \hat{N}_{j\bigcdot}}{\sqrt{\widehat{\widetilde{V}}_j}},
\end{equation}
where $N_j$ is defined in Eq \eqref{eqn:trueN}, $\hat{N}_{j\bigcdot}$ in Eq \eqref{eqn:N.hat.j}, and $\widehat{\widetilde{V}}_j$ in Eq  \eqref{eqn:V.hat.tilde.j}. The null and alternative hypotheses are $H_o: E[N_j - \hat{N}_{j\bigcdot}] = 0$ and $H_a: E[N_j - \hat{N}_{j\bigcdot}] \not= 0$. With a significance level of $\alpha$, the two-sided $(1-\alpha)\times 100\%$ CI for the estimated number of events is 
\begin{equation} \label{eqn:CI}
\hat{N}_{j\bigcdot} \pm z_{1-\frac{\alpha}{2}} \sqrt{\widehat{\widetilde{V}_j}} \quad.
\end{equation}
The two-sided alternative hypothesis and CI are used here because we want to flag not only when observed events are more than expected but also when the observed value is less than expected. Depending on the question of interest, this CI can be one-sided if only excessive or deficient events are wanted, or even asymmetrically two-sided if we want to be more sensitive about excessive events than deficient events. 

\subsection{Theoretical Justification}\label{sec:survival_theojust}
The three components of $\widehat{\widetilde{V}}_j$ are derived from the definitions of $\hat{N}_{jl}$ in Eq \eqref{eqn:N.hat.jM} and $\hat{N}_{j\bigcdot}$ in Eq \eqref{eqn:N.hat.j}. By its definition, the variability of $\hat{N}_{j\bigcdot}$ comes from the two plug-in estimators, $\hat\beta_{nl}$ estimated from the mixed effect AG model and $d\hat\Lambda_{nl}(s)$ in Eq \eqref{eqn:dlambda}. We decomposed the variability by hierarchical layers. First, the number of events for a level-2 group within a level-1 group follows a Poisson process and has an inherent variability from the model. The variance of a Poisson distribution is the event rate, which is estimated by $\hat{N}_{j\bigcdot}$. This component records the variability from the fixed effects. Going up one level, there is variability across level-1 groups, which are captured by the random effects, and we applied the block jackknife to estimate the covariance matrix. Lastly, $\hat{N}_{j\bigcdot}$ comes from taking the average over $M$ imputations and there is variation across the different MI copies. If no MI is involved, the third component can be omitted in Eq \eqref{eqn:V.hat.tilde.j}. The rest of the components in $\hat{N}_{j\bigcdot}$ are observed data that contribute to the variability through the parameters but have no variability on their own.

Although similar to $m$-fold cross-validation, where a fold corresponds to a block, the block jackknife specifies each block to have the same number of elements whereas cross-validation does not necessarily have folds of equal lengths. The block jackknife 
is computationally simpler than the non-parametric bootstrap including its alternatives, $m$ within $n$ bootstrap and subsampling. It has been shown that when properly normalized, the block jackknife estimator converges to an F-distribution at rate $n$, which indicates that it can obtain asymptotically valid confidence ellipses for the true parameter \citep{kosorok2008introduction}. The block jackknife has a hyperparameter, the number of blocks $m$, that could be used to adjust the bias-variance trade-off. The exploratory results in Sections \ref{sec:survival_sims} and \ref{sec:survival_results} will show that block jackknife gives reasonable, well-validated variance estimation. Besides block jackknife, we have looked into the bootstrap and jackknife methods to estimate the second component and their results are summarized in Appendix A. 

\section{Simulations}\label{sec:survival_sims}
\subsection{Simulation Settings} \label{sec:survival_sims_settings}
To evaluate the proposed risk-adjusted method, we conducted various simulations where parameter settings were chosen to mimic the clinical data used in Section \ref{sec:survival_clinicalimp}. In a hypothetical situation, assume there are $10,000$ subjects (i.e., level-2 groups, denoted by $i$) at $150$ health care programs (i.e., level-1 groups, denoted by $j$), and each subject has an equal probability of going to each program. We assumed a nested effect of the hierarchy where one subject belongs to one and only one program. 
Each patient was assumed to have 4 encounters per year which is the standard of care. We studied two time periods, 2012 to 2014 (3 years) and 2014 (1 year), and validated the predictions of number of events and their variation on year 2015. A total of $10$ covariates were defined with five time-invariant variables and five time-varying variables, all randomly generated from multivariate normal distributions. Detailed information for our simulation settings can be found in Appendix B, which includes specific definitions of time points, covariates, event times, and censoring times as well as a summary of simulation constants and parameters (Table \ref{tab:sim_params}). Censoring rates and event rates are determined to mimic the CF clinical data. The values of the baseline hazard and the censoring time parameter are determined by the censoring rate.

Since there were only $10$ covariates and no missing data, we skipped the MI and feature selection steps in the pipeline illustrated in Figure 1. 
After the event time, censoring time, and at-risk intervals were generated, we moved on to survival modeling. For each iteration, we started with fitting a Cox proportional hazards model. 
Taking into account correlated observations, robust standard errors were utilized by identifying level-1 (e.g., program) and level-2 (e.g., patient) as correlated groups. The R package \texttt{survival} was used for the Cox model. Next, the mixed effect survival model was applied to the same covariates using the initial values from the Cox model to help with convergence. The computation tool we used to fit the frailty model was the R package \texttt{coxme}, which assumes that the random effects follow a Gaussian distribution \citep{Rcoxme} and is deemed more efficient because of the use of semi-parametric estimation in the lognormal frailty model. Instead of treating random effects as missing data and applying the EM algorithm, which has been proven to be slow, \texttt{coxme} incorporated random effects by penalizing the partial likelihood that can be easily implemented by adding a penalty term to standard Cox semi-parametric models \citep{therneau2003penalized}. The rest of the pipeline (resampling, variance estimation, and validation) was followed as described in the diagram. The clinical use case in Section \ref{sec:survival_clinicalimp} will explore all steps in the pipeline including MI. All simulation calculations were performed in R 3.6.1 \citep{R2019}. 

\subsection{Simulation Results} \label{sec:survival_sims_results}

The 3-year training period was set to be from January 1, 2012 to December 31, 2014. The 1-year training period was from January 1, 2014 to December 31, 2014. The 1-year test period for both training periods was January 1, 2015 to December 31, 2015. For each of the three periods, 100 simulated datasets were generated following the assumptions and definitions in Appendix B. In the clinical data, the observed censoring rate is between $0.67$ and $0.83$ for 2012-2014 and between $0.86$ and $0.95$ for single years 2013, 2014, and 2015. The observed second event rate for the 2012-2014 period is between $0.0001$ and $0.005$. To mimic the CF data, the censoring rate in our $100$ simulated datasets has mean (standard deviation, SD)) $0.78(0.05)$ for the 3-year period and mean (SD) $0.90 (0.02)$ for the two 1-year periods. The second event rate has mean (SD) $0.004(0.003) $. All values are similar to the CF data used in Section \ref{sec:survival_clinicalimp}. 

In the first step of the method (Figure 1 
), all estimated coefficients are between $0.45$ and $0.53$ when the true parameter for all covariates is $0.5$, indicating that the Cox model and frailty model fit our training data well. Detailed results of the covariate coefficients, $\beta$'s, are presented and discussed in Appendix B. The number of events were estimated in the second step. To visually inspect the estimated number of events across simulations, we examined four histograms of the estimated versus the observed number of events across 100 simulated datasets (Figure 2 
). The four subplots are results trained and validated on 2012-2014 data (top left), trained on 2012-2014 and validated on 2015 (top right), trained and validated on 2014 (bottom left), trained on 2015 and validated on 2015 (bottom right), respectively. Although it is more important to have different training and validation data, we studied results from training and validating on the same data as a comparison reference. The estimated Spearman correlation coefficients between the estimated and observed number of events for each of the four subplots are $0.71, 0.24, 0.50, 0.37$, indicating weak to moderate positive correlations. Overall, all four distributions of the estimated events have roughly similar means and ranges as the four distributions of observed events, but the heights of the modes can vary. The top left histogram has the closest observed and estimated distributions because of the same training and validation data. The top right histogram is the second closest although the estimated distribution is denser near the mode than the observed distribution. The estimated and observed distributions differ more when the training data is 1 year instead of 3 years (the bottom two histograms), which can be explained by regression to the mean. 

One of the main contributions of our method is to apply block jackknife resampling to estimate the variance as well as the CIs in Eq \eqref{eqn:CI}. To observe results from various situations, three values of $m$ (the number of blocks in the block jackknife) were used and different confidence levels were explored spanning from $0.7$ to $0.995$. Table \ref{tab:sims_2014} contains the coverage of the estimated risk-adjusted CIs for the 3-year period and 1-year period. When trained on the 2012-2014 period and validated on the 2015 period (top right), the CI coverage is relatively close to the desired confidence level, the differences of which are within $3.3\%$ for all $\alpha$'s and $m$'s. In comparison, the coverage differences are all within $1.6\%$ for the overfitted 2012-2014 validation (top left). When trained on the 2014 period and validated on the 2015 period (bottom right), the CI coverage is relatively close to the desired confidence level but not as close as the 3-year period, with all differences within $6\%$ for all $\alpha$'s and $m$'s. In comparison, the largest absolute coverage differences are all within $0.5\%$ for all $\alpha$'s and $m$'s when trained and validated on the same 2014 period (bottom left). The 2015 validation of 2012-2014 training (top right) does not have a clear trend that higher confidence levels have more coverage or that certain values of $m$ constantly give higher coverage, but the differences in coverage among different $m$ values are closer when confidence level is larger (i.e., when $\alpha$ is smaller). For the 2015 validation of 2014 training (bottom right), however, the coverage is better as confidence level $1-\alpha$ increases, and $m=15$ seems to produce lower coverage differences compared with $m=5,10$ albeit not by very much. Overall, the 2015 validation results are better using 3-year data than 1-year data in terms of absolute coverage difference and this could be because 3-year data are richer and contain more heterogeneity information for the proposed model to learn, thus leading to less overfitting and better generalizability.

\section{Clinical Application}\label{sec:survival_clinicalimp}

Cystic fibrosis (CF) is a chronic, genetic disorder where defects in a chloride ion channel lead to excessively thick and sticky mucus throughout the respiratory and digestive systems. This impaired mucus clearance predisposes patients to respiratory infections and chronic lung damage, the main cause of morbidity and mortality in patients with CF. Two bacteria, methicillin-resistant \emph{Staphylococcus aureus} (MRSA) and \emph{Pseudomonas aeruginosa} (PA) are particularly harmful and lead to decreased lung function and poor outcomes for patients with CF. 
Patient-to-patient transmission within the healthcare system occurs in both the inpatient and outpatient settings and remains a concern for CF patients and healthcare programs\citep{saiman2003infection}. Monitoring incidence infection rates can provide a measure of potential transmission and these measures are important when assessing IP\&C policies. An important database for CF research in incidence tracking is the CFFPR, an ongoing study established to track survival trends and key patient outcomes, capturing longitudinal data of more than 95\% of the CF patients from over 200 accredited CF care programs in the U.S\cite{von2016exploring}. It has high participation rate and low loss to follow-up, but there is no centralized reporting for incident infections in the U.S. and the results are only shared to the programs at the end of each year \citep{knapp2016cystic}. As a result, CF programs currently do not have an accurate and timely way of tracking their incidence rates of MRSA and PA\citep{stoudemire2019cystic}. Furthermore, it is important to account for the characteristics of each program because they have a considerable variability in terms of IP\&C. To bridge this gap, we apply our proposed pipeline to the CFFPR data from 2012 to 2015 to learn, adjust, predict, and monitor the incidence of bacterial infection for each CF program.

\subsection{Preprocessing}\label{sec:survival_preprocess}

The MRSA infection is defined as having a positive respiratory culture for MRSA - a bacterium that has developed resistance to penicillin-based antibiotics. The PA infection is defined as having a positive respiratory culture to \emph{Pseudomonas aeruginosa}. Infections could occur again after a pre-specified washout period. Sensitivity analyses had been done previously to compare 2, 5, 10 years of a washout period and concluded that shorter look-back intervals did not alter overall estimates of incidence or changes in incidence \citep{salsgiver2016changing}. Consequently, the incidence rate is defined as the number of incident cases of a bacterium (with a 2-year look-back free of infection) divided by the at-risk population of the bacterial infection adjusted for length of being at-risk. Hence, the start date of an at-risk period for a patient is defined as the first day that the patient has had 2 years free of infections since the last infection. The end date is the date of event, lost to follow-up date, or the cutoff date (the last day of time period of interest).
We applied the following exclusion criteria to define the at-risk population. We excluded all encounters of patients 1) who did not have encounter dates and 2) whose start date of the at-risk day was the infection date. We excluded encounters, not the entire patient, 3) who had gaps in encounter data for more than 18 months (otherwise we assumed the patient's infection status stayed the same during the gap) and 4) after organ transplants. Babies less than two years old were considered at-risk even though they do not have two years of data to look back at. Relocation and change of program were allowed. After data cleaning, we constructed at-risk intervals consisting of start and end dates of being at-risk as well as a censoring indicator for each patient. Each patient can have 0, 1, or 2 infection(s) for a 3-year time period as there are at most two infections in three years given the two-year look-back. Each patient can have at most one infection for a time period of less than two years. 

The CFFPR data have a 2-level hierarchy where level-1 is each CF program and level-2 is CF patient who goes to a CF program. We combined several modalities of data (demographics and diagnosis, encounter, annualized, and program characteristics) by unique patient ID, program ID, and review year. Because these modalities have different levels of granularity, the time scales of the risk factors could not vary or vary by year or by day. The potential risk factors could be time-invariant (demographic data) or time-varying covariates (encounter and annualized data). The values of time-varying covariates were associated with the start dates of the corresponding at-risk intervals (as opposed to the end dates) as we assumed that the time-varying covariates did not change between the current encounter and the next consecutive encounter for gaps less than 18 months. Note that the start date of one at-risk interval may not be an actual encounter date because we derived the hypothetical at-risk start date by going back two years from an infection date. Each of the potential risk factors was chosen, preprocessed, and transformed (if necessary) by both the clinicians and biostatisticians on board. The reasons that we excluded some covariates include high correlation and multicollinearity, rare events, and more than 50\% missingness, and covariates clinically relevant. We studied six combinations of training and validation data for the period of 2012-2015: training 2012 to validate 2013, training 2013 to validate 2014, training 2014 to validate 2015, training 2012-2013 to validate 2014, training 2013-2014 to validate 2015, and training 2012-2014 to validate 2015. For a concise result presentation, we selected two exemplary combinations. We chose to present the year 2014 as the training set and year 2015 as the validation set because they are the most recent data. We will also present a training set of three years 2012-2014 with the same 2015 validation because it is the longest data period. Other combinations are omitted because results were more similar when the numbers of years in the training set were the same but more different when the number of years differed. The training and validation sets were preprocessed in the same way as described above. Only CF programs that appear in the training set were included in the validation set, and only the important risk factors in the training set were included in the validation set.

\subsection{Multiple Imputation}\label{sec:survival_MI}

Although the demographic and clinical data are fairly complete, CFFPR still has missing data (e.g., genotype, FEV1, mother's education) which might be associated with the epidemiology of the condition, lack of follow-up encounter, and socioeconomic status and gender\cite{mendelsohn2015characterization}. Different imputation methods were applied to each modality separately before merging. We generated $M=10$ copies of imputed datasets. \emph{Encounter data}: We applied the last observation carried forward (LOCF) and the next observation carried backward (NOCB) to variables in the encounter data.  Missing FEV1 (forced expiratory volume in one second) values were replaced by the max FEV1 value in the past 365 days for each patient. After LOCF and NOCB, the remaining incomplete variables were missing for all observations/encounters of the patients. Taking into account the longitudinal nature of the encounter data, we applied joint modeling imputation \citep{carpenter2012multiple} in the \texttt{mitml} R package \citep{Rmitml} with unique patient ID as the random effect. This is an MCMC imputation algorithm suited for multilevel data with continuous and categorical variables, which matches with the longitudinal nature of our encounter data. \emph{Annualized data}: We did not apply LOCF or NOCB because a year was considered too long to be carried forward or backward unlike the more frequent encounter data. Instead, joint modeling was applied with patient ID as a random effect. 
\emph{Demographic data}: We applied the multiple imputation by chained equations (MICE) method with random forests (RF) because there were no random effects in demographic data since it is one observation per unique subject and MICE can impute nominal variables
(e.g., mutation information such as F508 - the most common disease-causing mutation in CF).

\subsection{Survival Model and Variable Selection}\label{sec:survival_VS}

The random effects in the mixed effect model could be nested or crossed. 
Unlike the simulations, two crossed random effects (one for each hierarchy) were used in the clinical application as CF patients can relocate over the follow-up period and do not necessarily go to only one program. With more than 70 potential risk factors in the preprocessed data, we wanted to narrow them down and identify only the important ones that are associated with incidence rates of MRSA and PA. Step-down selection was used for this purpose. Step-down selection typically starts with modeling on all covariates and removing one insignificant covariate at a time until all covariates left have $p$-value less than the significance level. The remaining covariates are called important risk factors. Under several reasonable, generalized assumptions, meaningful recursive feature elimination methods with kernel machines can find the correct feasible feature space with uniform consistency   \citep{dasgupta2019feature}. 

We made two changes to the classic step-down selection procedure to better cater to our situation. First, removing variables one at a time for over 70 covariates would be time-consuming. We allowed the step-down selection to drop more than one variable for each iteration to speed up the variable selection process. More specifically, we removed three variables at a time in the early selection stage where the $p$-values of the dropped variables were greater than or equal to $0.5$, and drop two variables at a time when $p$-values were strictly between $0.25$ and $0.5$. When the p-values were less than or equal to $0.25$, we slowed down the elimination process and dropped one variable at a time. We stopped removing variables when the highest $p$-value was below $0.1$. Second, due to MI, we fit the two survival models to each imputation copy separately before pooling the estimates from all $M$ MI datasets together to determine which variable(s) to drop. As in the simulations, we used the Cox model first to acquire initial values for the coefficients before fitting the mixed effect AG model. The same R packages (\texttt{survival} and \texttt{coxme}) were applied to the CF data. After one iteration was done, we repeated the same procedure with newly dropped covariates until all $p$-values were below $0.1$. We pooled the coefficient estimates one last time and used the final results to determine important risk factors. This stepwise selection was performed separately for MRSA and PA since different infections do not necessarily have the same risk factors.  All data cleaning and analyses  proposed in the pipepline were performed in R 3.6.1 \citep{R2019}.

\subsection{Results}\label{sec:survival_results}

The number of observations varies by the time period. For 2012-2014 (the longest period we looked at), MRSA had 219,251 observations for 18,366 patients and PA had 123,341 observations for 13,228 patients after preprocessing. For 2014, MRSA had 70,363 observations for 15,810 patients and PA had 41,085 observations for 10,505 patients preprocessing. The number of covariates in the preprocessed data was 79 for MRSA and 72 for PA. After step-down selection of 2012-2014, 41 out of 79 (MRSA) and 45 out of 72 (PA) risk factors were selected. For the single year 2014, there were 19 out of 79 risk factors selected for MRSA and 34 out of 72 for PA. The overall mean number of yearly encounters per patient was around $5.6$. Important factors significantly associated with MRSA infections included season, region, number of hospitalization/outpatient visits in the past year, birth year, and the proportions of whether or not a bacterial culture was done, need-based insurance (Medicare/Medicaid/State level), feeding, smoking, etc. PA had additional important factors such as mutation class, Hispanic race, salt supplement, FEV1, in addition to season and region. There were $M=10$ multiple imputed copies and $m=5,10,15$ numbers of blocks, all were the same as in the simulations.

\begin{figure}
	\centering 
	\includegraphics[scale=0.35]{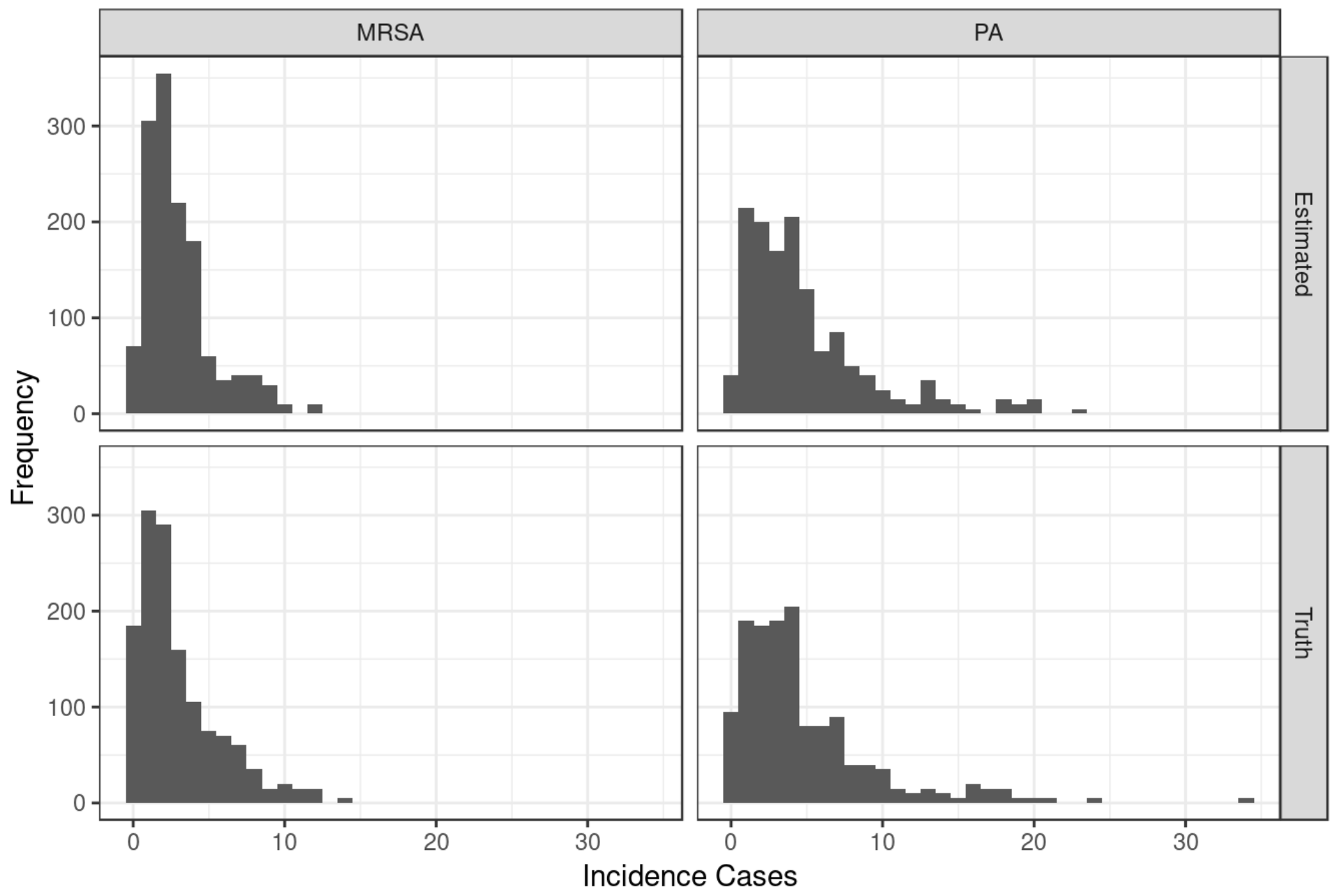}
	\includegraphics[scale=0.335]{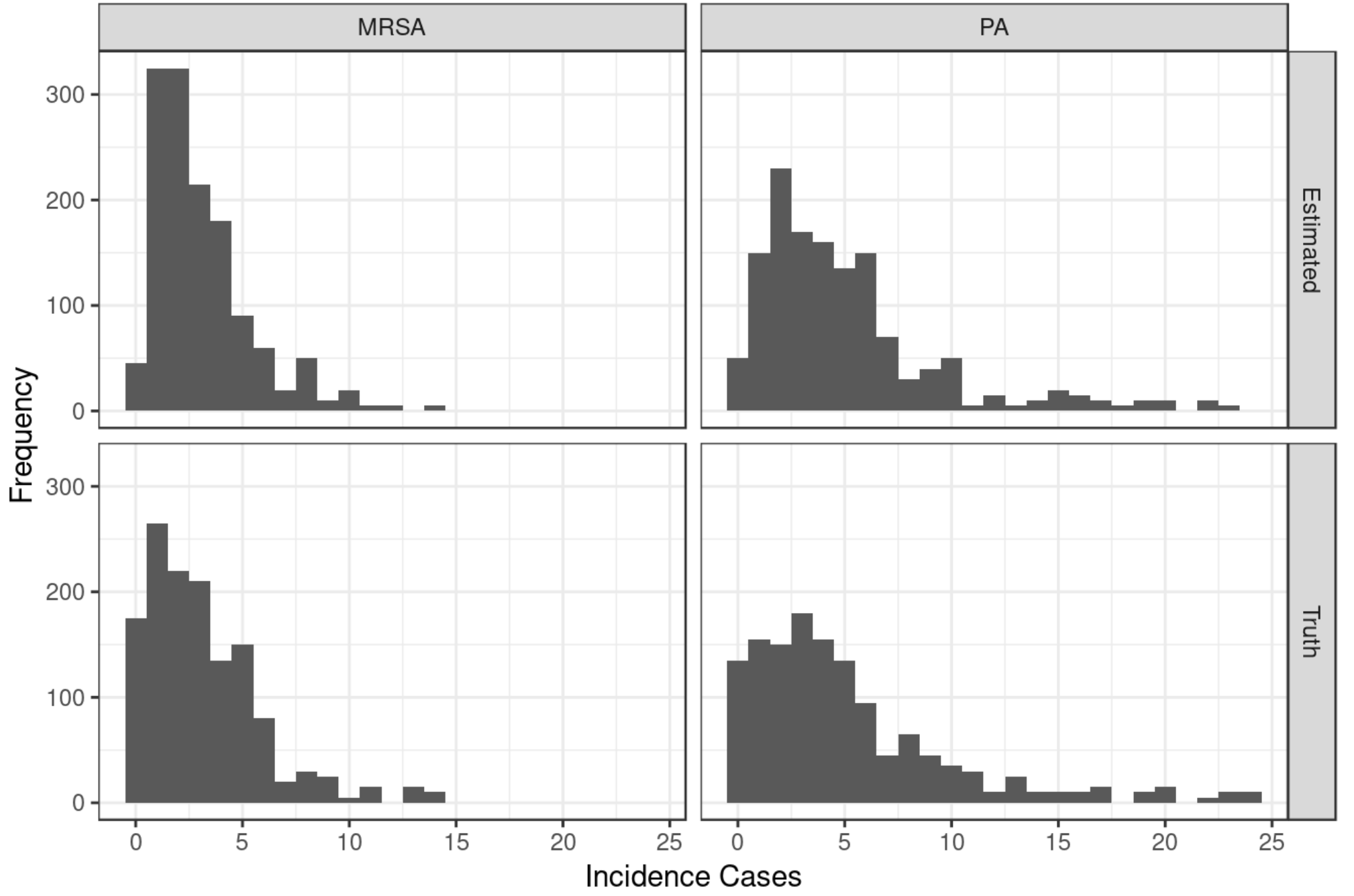}
	\caption{Histogram of estimated versus observed/true number of 2014 (top) and 2015 (bottom) MRSA and PA incidence cases where the estimated incidence cases come from a risk-adjusted model trained with 2014 data}
	\label{fig:2014_on_2015_point}
\end{figure}

\begin{figure}
    \centering
    \includegraphics[scale=0.5]{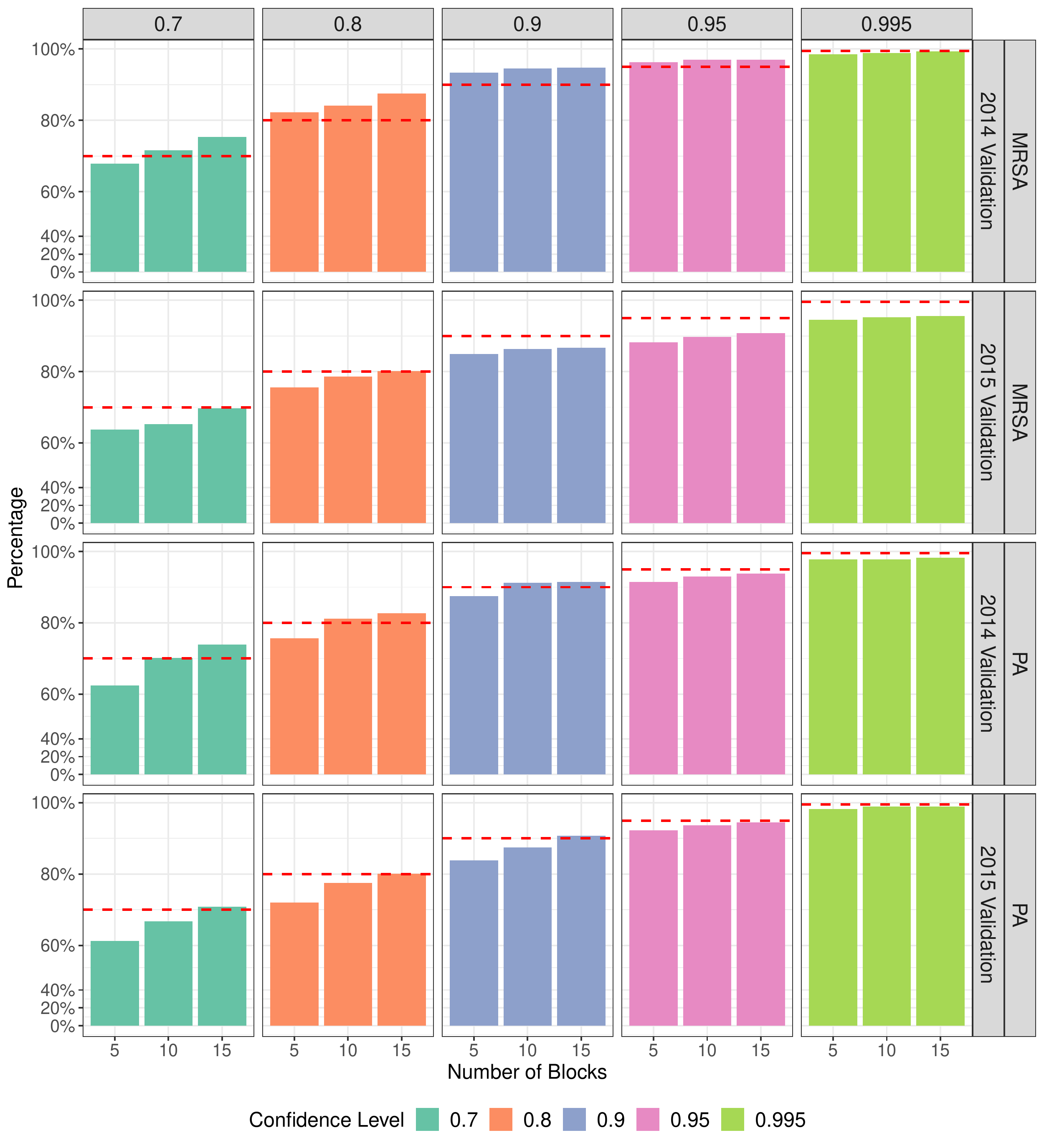}
    \caption{Coverage of risk-adjusted confidence intervals estimated from 2014 CFFPR data and validated on 2014 or 2015 for MRSA and PA across five confidence levels and three numbers of blocks. Coverage is the proportion of CF programs whose true number of 2014 or 2015 incidence cases is contained in the risk-adjusted confidence interval trained from 2014 data. The horizontal dashed lines mark the confidence levels for easier comparison. The 0\% to 50\% part of y-axis was folded by a factor of 4 for better presentation. }
    \label{fig:2014_on_2015_ci}
\end{figure}

\emph{Risk-Adjusted Incidence Modeling}.  Before validating with new data, we trained the risk-adjusted survival models with 2014 data and estimated the number of events for the same 2014 data. The top four plots in Figure \ref{fig:2014_on_2015_point} 
show that the distribution of the observed incidence cases was similar to the estimated incidence cases for both bacteria even at the tails. The correlation between the estimated and the observed incidence was strong as the estimated Spearman correlation coefficients were $0.68$ for MRSA and $0.77$ for PA. Based on the first and third rows of Figure \ref{fig:2014_on_2015_ci}
, almost all absolute coverage differences were less than $5\%$, implying that the coverage of the risk-adjusted confidence intervals was close to the desired confidence level for both bacteria overall. This is expected as the training and validation data are the same. For MRSA, $m=5$ had the closest coverage to true confidence level for middle confidence levels ($0.8$ to $0.95$). For PA, the coverage of $m=10$ was the closest for lower confidence levels ($0.7$ to $0.9$). The difference between coverage and true confidence level decreased as confidence levels increased ($\geq 0.95$) for all $m$ values and both bacteria. High confidence levels required wider confidence intervals and thus the bias-variance trade-off controlled by $m$ was not as prominent. A table with detailed numbers of the same CI coverage results is in Appendix C.

\emph{Validation with New Data}. We validated the model trained from 2014 data on the new 2015 data. The distributions of observed versus estimated incidence cases were similar (bottom four plots in Figure \ref{fig:2014_on_2015_point}) including the right-skewed tails. The estimated Spearman correlation coefficient between the estimated and the observed incidence was $0.59$ for MRSA and $0.79$ for PA (moderate to strong correlations). The second and fourth rows of Figure \ref{fig:2014_on_2015_ci} 
show that the validation coverage rates were a bit worse by a small amount compared with the results from 2014 training and 2014 validation, since the validation set was new data that the model had not seen. As confidence level went up, the coverage difference decreased for both MRSA and PA. The overall coverage of our CI was reasonably close to the desired confidence level as the coverage rates were all within $10\%$ away from the dashed line for all three $m$'s,  and all but one coverage level were less than $5\%$ away from the dashed line for $m=10,15$. Besides, $m=15$ gave the best coverage for both bacteria. To some extent, coverage results depended on the hyperparameter $m$ but were within a reasonable range in general. Overall, results showed that the block jackknife can estimate variabilities in the incidence cases well given that the coverage differences were reasonably small regardless of the value of $m$. 

\begin{figure}
    \centering
    \includegraphics[scale=0.5]{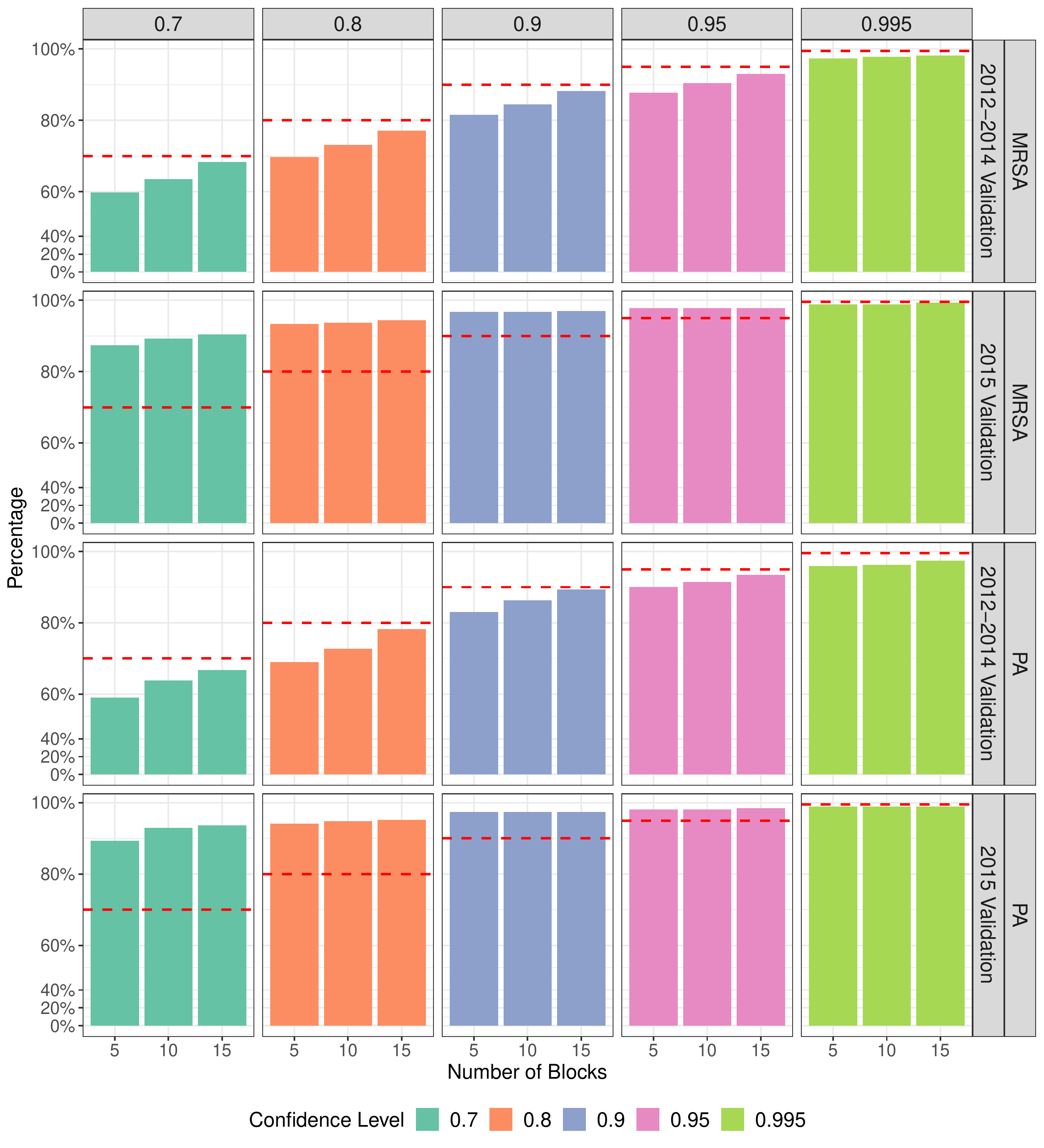}
    \caption{Coverage of risk-adjusted confidence intervals estimated from 2012-2014 CFFPR data and validated on 2012-2014 or 2015 for MRSA and PA across five confidence levels and three numbers of blocks. Coverage is the proportion of CF programs whose true number of 2012-2014 or 2015 incidence cases is contained in the risk-adjusted confidence interval trained from 2012-2014 data. The horizontal dashed lines mark the confidence levels for easier comparison. The 0\% to 50\% part of y-axis was folded by a factor of 4 for better presentation.}
    \label{fig:2012-2014_on_2015_ci}
\end{figure}

\emph{Multiple years of training data}. We picked the 2012-2014 period as the training set and validated on both the 2012-2014 and 2015 data. As Figure \ref{fig:2012-2014_on_2015_ci} 
shows, overcoverage was observed in the second and fourth rows (coverage rates were all higher than confidence levels) as the training set contained more than one year of CFFPR data, while the training coverage differences in the first and third rows stayed relatively small as before. Simulation results in Section \ref{sec:survival_sims_results} showed that three years of training data produced CIs with higher coverage than one year of data. Three years of data allow recurrent events and have a larger sample size but the data seem to be noisier and contain more complex situations in the real world. For instance, we assumed all patients visit their CF program four times a year in simulations per IP\&C recommendation, while this is not always the case in reality. Patients also might relocate over the years to a different program and nested random effects may no longer be valid. When the training set contains more than one year, more encounters with longer gaps are included which can increase the variability of data. The proposed model learned from this complexity and generated larger variance estimates and wider CIs to compensate for the variability which leads to overcoverage in the validation set. This is particularly noticeable when the confidence level was low ($0.7$ or $0.8$) for 2015 validation, compared with higher but more common confidence levels such as $0.9$ and above. Histograms of estimated versus observed numbers of 2012-2014 and 2015 incidence cases for both MRSA and PA as well as a table of the detailed results in Figure \ref{fig:2012-2014_on_2015_ci} 
can be found in Appendix C. Also contained in Appendix C are our investigation of the cohort effects across different periods in the CF patients as well as residual plots as another model performance metric in addition to the coverage metric.

\section{Discussion}\label{sec:survival_discuss}

A risk-adjusted model was developed to learn from right-censored hierarchical data with recurrent events and estimate the number of events with approximately valid confidence intervals. The variability of predicted incidence cases had three sources of variation, one of which was obtained with the block jackknife method. We carefully preprocessed the data and recommended useful tools such as multiple imputation and variable selection to produce clean and concise input data before modeling. Simulations were conducted to evaluate our methodology of the risk-adjusted confidence intervals. CF registry data were used as a clinical use case to evaluate the practicality of the proposed pipeline. Overall, our results showed promising effectiveness of the risk-adjusted incidence models in terms of coverage. Although increased sample size is generally beneficial, focusing primarily on data closer to the time of interest can be more predictive in our CF example. This implies that larger data are beneficial when clean and well-organized, but larger observational data often come with more complexity (e.g., due to logistical challenges of following a large number of people over a long period of time), which can increase the difficulty of variance estimation. Our explorations imply that data recency and data quality can be more important than data quantity. 

 The method can estimate the recurrent survival events and its variability for each program and package everything into a confidence interval which is easy to acquire and understand. Incidence estimates and their confidence intervals have been organized into a spreadsheet and the built-in formulae can help determine whether or not to flag the program if the observed incidence provided falls above the estimated confidence interval. Additional clinical results can be found in Stoudemire et al.\cite{stoudemire2020}, which also provides quarterly reports on MRSA and PA incidence for each program. The reports validated on the first 3, 6, 9, months when the validation year had not ended, so unusual incidence rates can be detected earlier and infection control procedures can be implemented sooner. 

This study has several limitations. First, the level-1 group, the highest level in the hierarchy, is usually a social entity or some group of subjects that can vary in size. For example, the size of CF programs in the U.S. could range from 10 to 300 patients. Our model did not preprocess small programs ($n < 50$) differently for two main reasons: i) feature selection did not select program size as an important risk factor and ii) our results implied that it is not always small programs whose risk-adjusted intervals fail to contain the observed number of incidence cases (especially when the training period is one year). However, accounting for small programs has potential benefits for improving the model performance and generalizability as smaller programs are prone to create outliers.  
More simulations need to be done to study the underlying intensity processes. Second, the general computation time for the entire pipeline varies and depends largely on the sample size. Many aspects of the analysis, such as bacterium, time period, and number of blocks, were computed in parallel, and most steps usually took a couple of hours even for the largest 2012-2014 data in this study. When the number of covariates is large (e.g., around 80 in the CF data), the computational bottleneck is likely to be in the stepdown selection step, where parallel programming cannot be utilized because current selection depends on prior selection. Further research can look into alternative feature selection methods that can improve computation time while being flexible with multiple imputation and various survival models.

Our study has several future directions. First, this study serves as the first step in the precision public health paradigm for infection prevention and control. Compared with precision medicine which finds the optimal treatment for each patient, precision public health focuses on recommendations tailored to large entities such as health institutions on a broader level \citep{kosorok2018precision, sperger2020}. Given that our proposed methods can provide contextual intelligence of bacterial infection rates for CF programs, a natural continuum is to identify modifiable program-level risk factors (frequency of bacterial culture performed, mask use, cleaning, CF knowledge education) and provide program-specific strategies to proactively prevent infections. Precision health is a new and growing area that is in need of methodology development. Second, our methods can be potentially extended to the area of multiple event types which can be a mixture of competing and non-competing risks. Taking the general health of elderly people as an example, hospitalization can be a recurrent event with death as a competing risk. In CF, clinicians might be interested in incidence rates at the same time as CF-related diabetes and liver diseases, which may or may not be competing with infections. Bivariate or multivariate survival models could be applied for multiple event types.

\section*{Acknowledgments}
This work was supported by the US Cystic Fibrosis Foundation Grant STOUDE18A0-D3 and the National Center for Advancing Translational Sciences (NCATS), National Institutes of Health, through Grant Award Number UL1TR002489. We would like to thank the CF Foundation (CFF) for the use of CFF Patient Registry data to conduct this. Additionally, we would like to thank the patients, care providers, and clinic coordinators at CF programs throughout the United States for their contributions to the CFF Patient Registry. The content is solely the responsibility of the authors and does not necessarily represent the official views of the NIH. The authors declare no potential conflicts of interest relevant to this article.







\begin{table}[ht!]
	\scriptsize
        \caption{Results of risk-adjusted model trained from 2012-2014 (top) and 2014 (bottom) simulated data validated on 2012-2014 and 2015 (top) and 2014 and 2015 (bottom) simulated data separately, where $1-\alpha$ is the confidence level, $m$ is the number of blocks in block jackknife estimation of variance, meanCoverage is the mean proportion of level-1 groups whose true number of validation cases is contained in the risk-adjusted $1-\alpha$ confidence interval estimated from the training data averaged across 100 simulations, and AbsCovDiff is the absolute difference between meanCoverage and $1-\alpha$ (closer to zero is more desirable)}
	\centering
	\begin{tabular}{rr||cc|cc}\toprule  
		$1-\alpha$ & m & meanCoverage & AbsCovDiff & meanCoverage & AbsCovDiff \\ \toprule 
		\multicolumn{2}{c||}{2012-2014 Training} & \multicolumn{2}{c|}{2012-2014 Validation} & \multicolumn{2}{c}{2015 Validation} \\ \hline
		0.7 & 5 & 0.706 & 0.006 & 0.724 & 0.024 \\ 
		& 10 & 0.711 & 0.011 & 0.729 & 0.029 \\ 
		& 15 & 0.716 & 0.016 & 0.732 & 0.032 \\ \hline 
		0.8 & 5 & 0.800 & 0.000 & 0.805 & 0.005 \\ 
		& 10 & 0.806 & 0.006 & 0.808 & 0.008 \\ 
		& 15 & 0.811 & 0.011 & 0.812 & 0.012 \\ \hline 
		0.9 & 5 & 0.895 & 0.005 & 0.882 & 0.018 \\ 
		& 10 & 0.901 & 0.001 & 0.884 & 0.016 \\ 
		& 15 & 0.905 & 0.005 & 0.886 & 0.014 \\ \hline 
		0.95 & 5 & 0.946 & 0.004 & 0.919 & 0.031 \\ 
		& 10 & 0.948 & 0.002 & 0.921 & 0.029 \\ 
		& 15 & 0.951 & 0.001 & 0.923 & 0.027 \\ \hline 
		0.995 & 5 & 0.991 & 0.001 & 0.962 & 0.033 \\ 
		& 10 & 0.992 & 0.002 & 0.964 & 0.031 \\ 
		& 15 & 0.992 & 0.002 & 0.964 & 0.031 \\ \hline 
	    \multicolumn{2}{c||}{2014 Training} & \multicolumn{2}{c|}{2014 Validation} & \multicolumn{2}{c}{2015 Validation} \\ \hline
		0.7 & 5 & 0.695 & 0.005 & 0.646 & 0.054 \\ 
		& 10 & 0.697 & 0.003 & 0.649 & 0.051 \\ 
		& 15 & 0.700 & 0.000 & 0.651 & 0.049 \\ \hline 
		0.8 & 5 & 0.802 & 0.002 & 0.742 & 0.058 \\ 
		& 10 & 0.803 & 0.003 & 0.745 & 0.055 \\ 
		& 15 & 0.806 & 0.006 & 0.747 & 0.053 \\ \hline 
		0.9 & 5 & 0.901 & 0.001 & 0.840 & 0.060 \\ 
		& 10 & 0.903 & 0.003 & 0.842 & 0.058 \\ 
		& 15 & 0.904 & 0.004 & 0.844 & 0.056 \\ \hline 
		0.95 & 5 & 0.951 & 0.001 & 0.895 & 0.055 \\ 
		& 10 & 0.952 & 0.002 & 0.897 & 0.053 \\ 
		& 15 & 0.953 & 0.003 & 0.898 & 0.052 \\ \hline 
		0.995 & 5 & 0.992 & 0.002 & 0.967 & 0.028 \\ 
		& 10 & 0.992 & 0.002 & 0.968 & 0.027 \\ 
		& 15 & 0.992 & 0.002 & 0.968 & 0.027 \\ \bottomrule  
	\end{tabular}
	\label{tab:sims_2014}
\end{table}

\clearpage

\section*{Appendix A: \\ Additional Theoretical Justification of Using Block Jackknife}\label{supp3:theo_justify}

The across-group variance $\hat{V}_{j\bigcdot}$ in Section 2.3 is tricky to estimate because, to the best of our knowledge, there are no known studies on the appropriate algorithm to use. In addition to the block jackknife, we explored two other methods, bootstrap and jackknife. We implemented two types of bootstrapping, one was resampling with replacement and one was the weighted bootstrap. When resampling level-1 groups as a whole (meaning all subsequent level-2 groups) with replacement, we encountered singularity issues. This error can be due to the fact that some level-1 groups were not included in the resampled data and attached the original dataset $D$ to the bootstrap sample $D^{bs}$ to make sure every level-1 group was included. We encountered similar errors using the combined data $\{D, D^{bs}\}$. Through careful checks we found out that the issue was because repeating the data produces tied failures times. 
This lead us to the idea of weighted bootstrap where
we did not resample with replacement but applied a weight randomly drawn from the exponential distribution with parameter $1$ for each level-1 group. The weighted bootstrap is more applicable in general, even when some nuisance parameters are not $\sqrt{n}-$consistent \citep{kosorok2008introduction}. However, we ran into optimization errors despite the different optimization methods we tried: Nelder-Mead \citep{nelder1965simplex}, BFGS \citep{broyden1970convergence, fletcher1970new, goldfarb1970family, shanno1970conditioning}, or Conjugate Gradient (CG) \citep{fletcher1964function}. An alternative to the bootstrap is the jackknife, also known as leave-one-out cross validation (LOOCV). We applied the jackknife to  $\hat\beta_{nl}^{(-v)}$ where we took out the $v$th level-1 grouping variable at a time. This gives us $$d\hat\Lambda_{nl}^{(-v)}(s) = \frac{\sum_{j=1, j \not= v}^N \sum_{i=1}^{c_j} \sum_{k=1}^{m_{ij}} d N_{jik}(s)}{\sum_{j=1, j \not= v}^N \sum_{i=1}^{c_j} \sum_{k=1}^{m_{ij}} 1\{L_{jik} < s \leq U_{jik}\} e^{Z_{jikl}(s)\hat\beta^{(-v)}_{nl} }}$$ and 
$$\hat{N}_{jl}^{(-v)} = \sum_{i=1}^{c_j}\sum_{k=1}^{m_{ij}} \int_s 1\{L_{jik} < s \leq U_{jik}\} e^{(\hat\beta_{nl}^{(-v)})^T Z_{jikl}(s)} d\hat\Lambda_{nl}^{(-v)}(s)$$ where $s$ represents all unique event time points in the original dataset and $l$ represents multiple imputation. Hence, $$ \hat{V}_{jl} = \hat{V}(\hat{N}_{jl}) = (N-1) \sum_{v = 1}^N \l(\hat{N}_{jl}^{(-v)} - \hat{N}_{jl} \r)^{\bigotimes 2}.$$ 
We did not run into any optimization or singularity issues with the jackknife method but the variance was large and some lower bounds of confidence intervals were far below zero, implying that it was not likely to give meaningful confidence intervals. This was not unexpected since the jackknife improves on bias by using as much training data as possible compared with other cross validation methods, but this improvement is achieved at the sacrifice of precision. This prompted us to consider the block jackknife, the middle ground, where we removed one block of level-1 groups at a time instead of only one level-1 group. 

\section*{Appendix B: Additional Information for Simulations}\label{supp3:sims}

\subsection*{Simulation Settings}
For both 3-year and 1-year periods, we used $g$ to denote encounters. Based on CF regulations, patients are expected to go to their CF programs about four times a year, so we created 12 encounter time points ($t_g$, where $g \in \mathcal{G} = \{1,\ldots,12\}$) for the 3-year training period and four encounter time points ($t_g$, where $g \in \mathcal{G} = \{1, 2, 3, 4\}$) for the 1-year period and assume all subjects in a simulated dataset had the same time points for each simulated dataset. The $t_g$'s are ordered in an increasing order with $t_1 = 0$. The rest of the $g-1$ time points were randomly generated from a uniform distribution $U(1,1100)$ or $U(1, 400)$ across the 3-year or 1-year period, all rounded to the nearest integer to imitate the number of days.

A total of 10 covariates were generated. The five time-invariant covariates, denoted as $\Z_{11}, \ldots, \Z_{15}$, followed a multivariate normal distribution
$MVN\l(\mu_1, \Sigma_1\r)$ with \\
$\mu_1 = \begin{pmatrix} 0, 0, 0 , 0, 0 \end{pmatrix}^T$ and 
$$\Sigma_1 = 
\begin{pmatrix} 
0.1 & 0.02 & 0.02 & 0.02 & 0.02 \\ 
0.02 & 0.1 & 0.02 & 0.02 & 0.02 \\ 
0.02 & 0.02 & 0.1 & 0.02 & 0.02 \\
0.02 & 0.02 & 0.02 & 0.1 & 0.02 \\
0.02 & 0.02 & 0.02 & 0.02 & 0.1 
\end{pmatrix}.$$ 
Here, we assume all $\Z_1$ variables were correlated weakly. The five time-variant covariates were denoted as $\Z_{21}(t), \ldots, \Z_{25}(t)$, independent and identically distributed, so each $\Z_{2p}(t)$ for $p=1,\ldots,5$ followed a multivariate normal distribution $MVN\l(\mu_2, \Sigma_2\r)$ across $t$. The mean vector is $\mu_2^{1} = \begin{pmatrix} -5, -4, -3, \ldots, 4, 5, 6 \end{pmatrix}^T$ for 12 time points or $\mu_2^{2} = \begin{pmatrix}  -1, 0, 1, 2 \end{pmatrix}^T$ for four time points. The covariance matrix is $$\Sigma_2^1 = a \begin{pmatrix}
12 &  11 & 10 & \ldots & 2 & 1 \\ 
11 & 12 & 11 &  \ldots & 3 & 2 \\
&  &  &  \ddots \\ 
1 & 2 & 3 & \ldots & 11 & 12 
\end{pmatrix} \text{  or  } \Sigma_2^2 = a \begin{pmatrix}
4 & 3 & 2 & 1 \\ 
3 & 4 & 3 & 2 \\
2 & 3 & 4 & 3 \\ 
1 & 2 & 3 & 4 
\end{pmatrix} $$ for 3-year and 1-year period respectively, where $a$ is a scalar randomly sampled from the uniform distribution $U(0, 0.1)$ for each simulation dataset. The covariance matrices were chosen to reflect correlation over time. 

The piece-wise survival time $S(t_g)$ followed a piece-wise exponential distribution with rate being the hazard: 
$$h_{ij}(t_g) = h_0 \exp(\Z_{1ij} \beta_1 + \Z_{2ij} (t_g) \beta_2 + b_{1i} + b_{2j}),$$ 
where $h_0$ is a constant baseline hazard function, $\beta_1, \beta_2$ are coefficients for the time-invariant and time-variant covariate $\Z_{1ij}, \Z_{2ij}(t_g)$ respectively, and $b_{1i}, b_{2j}$ are the random intercept effects on level-2 (patient level) and level-1 (program level). We assume no random slope effects for simplicity. Under the assumption that $\Z_{2ij}(t_g)$ does not change in the time interval $[t_g, t_{g+1})$ and the memoryless property of the exponential distribution, the event time is $T_{ij}^1 = t_g + S_{ij}(t_g)$ if $t_g + S_{ij}(t_g) < t_{g+1}$. Otherwise, there is no event between $t_g$ and $t_{g+1}$ and we simulated the next piece-wise survival time $S_{ij}(t_{g+1})$ from $h_{ij}(t_{g+1})$ and repeated the process until we reached the last time point. Let $W_{ij}$ denote the length of the pre-specified washout period for the $i$th patient at the $j$th program. The washout period is the time period during which new infections are not considered as incident cases. For simplicity, we assume $W_{ij}$ is constant for all $i,j$'s. For recurrent events, the second event time is $T_{ij}^2 = T_{ij}^1 + W_{ij} + S_{ij}(t_{g'}) $ if $T_{ij}^1 + W_{ij} + S_{ij}(t_{g'}) < t_{g'+1}$ where $g' = \{g \in \mathcal{G}: \max(t_{g'} \leq T_{ij}^1 + W_{ij})\}$. Otherwise, there is no event between $T^1_{ij} + W_{ij}$ and $t_{g'+1}$. We simulated the next piece-wise survival time $S_{ij}(t_{g'+1})$ accordingly and determined the second event time as $T_{ij}^2 = t_{g'+1} + S_{ij}(t_{g'+1})$ if $ t_{g'+1} + S_{ij}(t_{g'+1}) < t_{g'+2}$. We repeated this process the same way as the first event time described above. For all $i,j$
s, the independent censoring time for the $i$th level-1 group and $j$th level-2 group $C_{ij}$ is exponentially distributed with rate $1/600$ and $1/300$ for the 3-year and 1-year period. The event indicator is $\delta_{ij} = 1\{T_{ij} \leq C_{ij}\}$, where $1$ indicates an event and $0$ indicates censoring. The observed time is $Y_{ij} = \min(T_{ij}, C_{ij})$. There could be more than one $\delta_{ij} = 1$ because of recurrent events.

Table \ref{tab:sim_params} 
is a summary of all constants and parameters in the simulation introduced so far. Because we assume a nested structure between the level-1 group and level-2 group, all notations with subscript $ij$ can be written with subscript $i$ only, as $j$ is deterministic given $i$. To be more general, we still use $ij$ as it can also represent non-nested structures (e.g., crossed) between level-1 and level-2 groups.

\begin{table}[htbp]
	\centering 
	\caption{Simulation constants and parameters}
	\begin{tabular}{lccccc}\toprule 
		Parameter & Notation & & \multicolumn{2}{c}{Time Period} \\ \cline{4-5}
		& & & 3-year  & 1-year  & \\ \toprule 
		No. of groups, level-1 & $n_2$ && \multicolumn{2}{c}{150} \\ \hline 
		No. of groups, level-2 & $n_1$ && \multicolumn{2}{c}{10,000} \\ \hline 
		No. of time points & $|\mc{G}|$ && $12$ & $4$\\ \hline 
		Length of study period & && 1100 & 400 \\  \hline 
		Encounter time & $t_g, g \in \mc{G}$ && $U(12, 1100)$ &  $U(4, 400)$ \\ \hline 
		No. of time-invariant covariates & $p_1$ &&  \multicolumn{2}{c}{5} \\ \hline 
		No. of time-variant covariates & $p_2$ &&  \multicolumn{2}{c}{5} \\ \hline 
		Time-invariant covariate & $\Z_1$ && \multicolumn{2}{c}{$MVN(\mu_1, \Sigma_1)$}\\ \hline 
		Time-variant covariate \\
		$\quad$ (i.i.d. across $p = 1, \ldots, p_2$) & $\Z_{2p}(t)$ && $MVN(\mu_2^1, \Sigma_2^1)$ & $MVN(\mu_2^2, \Sigma_2^2)$ \\  \hline 
		Baseline hazard & $h_0$ && \multicolumn{2}{c}{$e^{-8}$} \\ \hline 
		Covariate coefficient & $\beta_1, \beta_2$ && \multicolumn{2}{c}{all 0.5} \\ \hline 
		Random intercept effect, level-1 & $b_j$ && \multicolumn{2}{c}{$N(0, G_j), G_j \sim N(0, 0.001)$} \\\hline 
		Random intercept effect, level-2 & $b_i$ && \multicolumn{2}{c}{$N(0, G_i), G_i \sim N(0, 0.001)$} \\ \hline 
		Piece-wise hazard function & $h_{ij}(t)$ &&  \multicolumn{2}{c}{$h_0 \exp(\Z_{1ij} \beta_1 + \Z_{2ij} (t) \beta_2 + b_{1i} + b_{2j})$}\\ \hline 
		Piece-wise survival time & $S_{ij}(t)$ && \multicolumn{2}{c}{$\exp(h_{ij}(t))$} \\ \hline
		Censoring time & $C_{ij}$ && $\exp(1/600)$ & $\exp(1/300)$ \\ \hline 
		Washout period & $W_{ij}$ && \multicolumn{2}{c}{730 (2 years)}\\ \hline 
		No. of Simulations & & & \multicolumn{2}{c}{100} \\	\bottomrule 
	\end{tabular}
	\label{tab:sim_params}
\end{table}

\subsection*{Additional Simulation Results}

\begin{table}[htb]
	\centering
	\caption{Means (SDs) of estimated covariate coefficients for three time periods based on the Cox proportional hazards model and mixed effects AG model across 100 simulations}
	\begin{tabular}{cccc}\toprule  
		Covariate & 2012-2014 & 2014 & 2015 \\ \toprule 
		\multicolumn{4}{c}{Cox Proportional Hazards Model} \\ \midrule  
		$\Z_{11}$	& 0.46 (0.07) & 0.51 (0.11) & 0.51 (0.11)\\
		$\Z_{12}$	& 0.47 (0.08) & 0.51 (0.11) & 0.51 (0.11)\\
		$\Z_{13}$	& 0.46 (0.07) & 0.49 (0.11) & 0.49 (0.11)\\
		$\Z_{14}$	& 0.45 (0.08) & 0.51 (0.12) & 0.50 (0.12)\\
		$\Z_{15}$	& 0.46 (0.08) & 0.49 (0.10) & 0.49 (0.10)\\ 
		$\Z_{21}$	& 0.46 (0.09) & 0.49 (0.12) & 0.50 (0.14)\\
		$\Z_{22}$	& 0.47 (0.05) & 0.50 (0.11) & 0.51 (0.12)\\
		$\Z_{23}$	& 0.47 (0.09) & 0.50 (0.16) & 0.49 (0.15)\\
		$\Z_{24}$	& 0.47 (0.10) & 0.53 (0.16) & 0.52 (0.16)\\
		$\Z_{25}$	& 0.47 (0.04) & 0.48 (0.14) & 0.48 (0.14)\\\midrule  
		\multicolumn{4}{c}{Mixed Effects AG Model} \\ \midrule 
		$\Z_{11}$	& 0.47 (0.07)	& 0.51 (0.11) &	0.52 (0.11) \\
		$\Z_{12}$	& 0.48 (0.08)	& 0.52 (0.11) &	0.51 (0.11) \\
		$\Z_{13}$	& 0.47 (0.07)	& 0.49 (0.11) &	0.50 (0.11) \\
		$\Z_{14}$	& 0.46 (0.07)	& 0.52 (0.12) &	0.50 (0.12) \\
		$\Z_{15}$	& 0.47 (0.07)	& 0.49 (0.10) &	0.49 (0.10) \\ 
		$\Z_{21}$	& 0.47 (0.09)	& 0.49 (0.12) &	0.50 (0.14) \\
		$\Z_{22}$	& 0.47 (0.05)	& 0.51 (0.12) &	0.51 (0.12) \\
		$\Z_{23}$	& 0.48 (0.09)	& 0.50 (0.16) &	0.50 (0.15) \\
		$\Z_{24}$	& 0.48 (0.09)	& 0.53 (0.16) &	0.52 (0.16) \\
		$\Z_{25}$	& 0.48 (0.04)	& 0.48 (0.15) &	0.48 (0.14) \\ \bottomrule  
	\end{tabular}
	\label{tab:survival_sim_coefs}
\end{table}

The estimated coefficients for $\Z_1$ and $\Z_2(t)$ with SDs are displayed in Table \ref{tab:survival_sim_coefs} 
using a Cox proportional hazards model and a mixed effect Andersen-Gill (AG) model. Overall, all estimated coefficients were reasonably close to the true value of $0.5$ regardless of the time-invariant or time-variant covariate or survival model. The two 1-year periods had smaller biases than the 3-year period but higher SDs. The 2012-2014 period had estimates lower than $0.5$ and we speculate this is due to the interference of recurrent events. The coefficients in the mixed effect model had smaller biases for the corresponding coefficients in the Cox model, indicating that adjusting for the random effects helped with estimation and it learned the structure in the data better. This confirms that our simulated data and models were generated and fit well. All covariates had p-values less than 0.05, implying that the coefficients were significantly different from $0$, which is expected as all covariates were involved in the definitions of the hazard function and survival time. 

\section*{Appendix C: Additional Information on Clinical Application}\label{supp3:clinical_app}

\subsection*{Patient and Program Characteristics by Year}\label{supp3:cohort}

In the clinical application, we used several years of CFFPR data to study the incident events. Here we investigate whether there were cohort effects in our data, i.e., whether patient and program characteristics changed over time. Since our analysis is mainly cross-sectional in terms of different time periods, cohort effect could lead to misleading results because it is difficult to discern whether results are due to the effects of risk factors or variations in the time cohorts studied. Tables \ref{supp3:cohort_MRSA} 
and \ref{supp3:cohort_PA} 
display descriptive statistics of selected variables for both bacteria (MRSA and PA) and four years (2012 to 2015). The variables selected are commonly used and easy to acquire for all years. Since each patient could have multiple encounters we take the most frequent value if the variable is categorical or take the mean value if the variable is continuous. This converts the original data to the patient-level data, which are then used to calculate the descriptive statistics across CF programs. The values in the tables are counts (percentages) for categorical variables and means (SDs) for continuous variables.  By comparing the summary statistics across years, we see that the program and patient characteristics stay reasonably stable for all four years. The only relatively large change is in program type, where we see more patients who go to adult programs than pediatric programs for both bacteria. This is not surprising as our patients grew older and some could be switched to adult programs. It is also reassuring to see birth year increased by one for each new year. Based on this comparison across years, we do not think the cohort effect was an issue for our analysis.

\begin{table}[htpb!]
    \centering
    \caption{Descriptive statistics of selected variables for  the MRSA data (counts (percentages) for categorical variables and means (SDs) for continuous variables)}
    \begin{tabular}{lrrrr} 
        \toprule 
        Variable & 2012 & 2013 & 2014 & 2015\\ \toprule 
        Center region \\ 
        \ \emph{Midwest} & 4102 (26.9) & 4069 (26.9) & 4281 (27.1) & 4384 (26.9) \\
        \ \emph{Northeast} & 3167 (20.8) & 3143 (20.8) & 3196 (20.2) & 3296 (20.2) \\
        \ \emph{South} & 4736 (31.0) & 4623 (30.6) & 4977 (31.5) & 5119 (31.4) \\  
        \ \emph{West} & 3255 (21.3) & 3293 (21.8) & 3356 (21.2) & 3510 (21.5) \\ 
        Center Type \\ 
        \ \emph{Adult} & 4515 (29.6) & 4902 (32.4) & 5676 (35.9) & 5985 (36.7) \\ 
        \ \emph{Pediatric} & 8948 (58.6) & 8466 (56.0) & 8353 (52.8) & 8479 (52.0)\\ 
        Start season & \\
        \ \emph{Spring} & 4414 (28.9) & 4325 (28.6) & 4562 (28.9) & 4593 (28.2) \\
        \ \emph{Summer} & 7462 (48.9) & 7530 (49.8) & 7910 (50.0) & 8263 (50.7) \\ 
        \ \emph{Fall} & 1457 (9.5) & 1426 (9.4) & 1533 (9.7) & 1580 (9.7)\\ 
        \ \emph{Winter} & 1927 (12.6) & 1847 (12.2) & 1805 (11.4) & 1873 (11.5) \\ 
        Mean program size & 153 (80.2) &  149 (80.9) & 146 (81.4) & 149 (83.6) \\
        Birth year & 1993 (13.5) & 1994 (13.8) & 1995 (14.1) & 1996 (14.2)\\ 
        Proportion of \\ 
        \ private insurance & 9705 (63.6) & 9555 (63.2) & 9891 (62.6) & 10096 (61.9) \\
        Proportion of \\ 
        \ medicaid insurance & 5244 (34.4) & 5095 (33.7) & 5633 (35.6) & 6017 (36.9) \\ 
        Proportion of \\ 
        \ other insurance & 2539 (16.6) & 2432 (16.1) & 2504 (15.8) & 2364 (14.5)\\ 
        Mean highest sweat value & 97.6 (21.2) & 97.3 (21.3) & 96.9 (21.5) & 96.4 (21.7) \\ 
        Mean number of  \\ 
        \ hospitalization last year & 0.61 (1.1) & 0.62 (1.1) & 0.61 (1.1) & 0.62 (1.2) \\ 
        Mean number of \\ 
        \ cultures last year & 3.7 (2.0) & 3.7 (2.0) & 3.8 (2.0) & 3.8 (2.0)\\ 
        \bottomrule 
    \end{tabular}
    \label{supp3:cohort_MRSA}
\end{table}

\begin{table}[htpb!]
    \centering
    \caption{Descriptive statistics of selected variables for the PA data (counts (percentages) for categorical variables and means (SDs) for continuous variables)}
    \begin{tabular}{lrrrr} 
        \toprule 
        Variable & 2012 & 2013 & 2014 & 2015\\ \toprule 
        Center Region \\ 
        \ \emph{Midwest} & 2695 (27.8) & 2667 (27.3) & 2947 (28.1) & 2984 (37.3) \\
        \ \emph{Northeast} & 1972 (20.3) & 2022 (20.7) & 2125 (20.2) & 2250 (20.5) \\  
        \ \emph{South} & 3115 (32.1) & 3139 (32.1) & 3389 (32.3) & 3588 (32.8)\\ 
        \ \emph{West} & 1918 (19.8) & 1958 (20.0) & 2044 (19.5) & 2128 (19.4) \\ 
        Center type \\ 
        \ \emph{Adult} & 1553 (16.0) & 1760 (18.0) & 2219 (21.1) & 2415 (22.1) \\ 
        \ \emph{Pediatric} & 6892 (71.1) & 6737 (68.8) & 6977 (66.4) & 7165 (65.4) \\ 
        Start Season \\
        \ \emph{Spring} & 3108 (32.0) & 3282 (33.5) & 3605 (34.3) & 3706 (33.8) \\
        \ \emph{Summer} & 1716 (17.7) & 1715 (17.5) & 1928 (18.4) & 2099 (19.2) \\ 
        \ \emph{Fall} & 853 (8.8) & 890 (9.1) & 973 (9.3) & 1013 (9.3) \\ 
        \ \emph{Winter} &  4023 (41.5) & 3899 (39.8) & 3999 (38.1) & 4132 (37.7)\\ 
        Mean program size & 155 (81.3) & 150 (80.8) & 146 (81.5) & 148 (83.1) \\
        Birth year & 1998 (11.4) & 1999 (11.6) & 1999 (12.0) & 2000 (12.2) \\ 
        Proportion of \\ 
        \ private insurance & 5854 (60.4) & 5964 (60.9) & 6311 (60.1) & 6544 (59.8) \\
        Proportion of \\ 
        \ medicaid insurance & 3786 (39.0) & 3698 (37.8) & 4209 (40.1) & 4506 (41.2) \\ 
        Proportion of \\ 
        \ other insurance & 1709 (17.6) & 1597 (16.3) & 1706 (16.2) & 1662 (15.2) \\ 
        Mean highest sweat value & 95.4 (22.0) & 95.2 (22.3) & 94.8 (22.6) & 94.3 (22.8)\\ 
        Mean number of  \\ 
        \ hospitalization last year & 0.50 (1.0) & 0.50 (1.1) & 0.51 (1.1) & 0.52 (1.1)\\ 
        Mean number of \\ 
        \ cultures last year & 3.66 (1.9) & 3.69 (1.9) & 3.75 (2.0) & 3.83 (1.9) \\ 
        \bottomrule 
    \end{tabular}
    \label{supp3:cohort_PA}
\end{table}

\clearpage
\subsection*{Additional Clinical Application Results}\label{supp3:clin_results}

The coverage of risk-adjusted confidence intervals trained from both 2014 and 2012-2014 data has been illustrated in Figures 4 and 5. Here, we provide the results in numbers in Tables \ref{tab:2014_on_2015} and \ref{tab:2012-2014_on_2015_mrsa}
as supplemental information. In addition to coverage, the absolute difference between the coverage and confidence interval is also presented in the tables for easier comparison.

\begin{table}[htbp!]
    \scriptsize
	\caption{Results of risk-adjusted incidence model trained from 2014 CFFPR data validated on 2014 (top) and 2015 (bottom) data, where $\alpha$ is the significance level, $m$ is the number of blocks in block jackknife estimation of variance, Coverage is the proportion of programs whose true number of 2014 and 2015 incidence cases is contained in the risk-adjusted $1-\alpha$ confidence interval trained on 2014 data, and AbsCovDiff is the absolute difference between the coverage and $1-\alpha$.}
	\centering
	\begin{tabular}{rr||cc|cc}\toprule  
		& & \multicolumn{2}{c|}{PA} & \multicolumn{2}{c}{MRSA} \\ \toprule
		$1-\alpha$ & m & Coverage & AbsCovDiff & Coverage & AbsCovDiff \\ \hline 
		\multicolumn{6}{c}{2014 Training, 2014 Validation} \\ \hline 
		0.7 & 5 & 0.625 & 0.075 & 0.679 & 0.021 \\ 
		& 10 & 0.702 & 0.002 & 0.716 & 0.016 \\ 
		& 15 & 0.739 & 0.039 & 0.753 & 0.053 \\ \hline 
		0.8 & 5 & 0.757 & 0.043 & 0.823 & 0.023 \\ 
		& 10 & 0.812 & 0.012 & 0.841 & 0.041 \\ 
		& 15 & 0.827 & 0.027 & 0.875 & 0.075\\ \hline 
		0.9 & 5 & 0.875 & 0.025 & 0.934 & 0.034 \\ 
		& 10 & 0.912 & 0.012 & 0.945 & 0.045 \\ 
		& 15 & 0.915 & 0.015 & 0.948 & 0.048 \\ \hline 
		0.95 & 5 & 0.915 & 0.035 & 0.963 & 0.013 \\ 
		& 10 & 0.930 & 0.020 & 0.970 & 0.020 \\ 
		& 15 & 0.938 & 0.012 & 0.970 & 0.020 \\ \hline 
		0.995 & 5 & 0.978 & 0.017 & 0.985 & 0.010 \\ 
		& 10 & 0.978 & 0.017 & 0.989 & 0.006 \\ 
		& 15 & 0.982 & 0.013 & 0.993 & 0.002 \\ \hline 
		\multicolumn{6}{c}{2014 Training, 2015 Validation} \\ \hline
		0.7 & 5 & 0.613 & 0.087 & 0.638 & 0.062 \\ 
		& 10 & 0.668 & 0.032 & 0.653 & 0.047 \\ 
		& 15 & 0.708 & 0.008 & 0.697 & 0.003 \\ \hline 
		0.8 & 5 & 0.720 & 0.080 & 0.756 & 0.044 \\ 
		& 10 & 0.775 & 0.025 & 0.786 & 0.014 \\ 
		& 15 & 0.801 & 0.001 & 0.801 & 0.001\\ \hline 
		0.9 & 5 & 0.838 & 0.062 & 0.849 & 0.051 \\ 
		& 10 & 0.875 & 0.025 & 0.863 & 0.037 \\ 
		& 15 & 0.908 & 0.008 & 0.867 & 0.033 \\ \hline 
		0.95 & 5 & 0.923 & 0.027 & 0.882 & 0.068 \\ 
		& 10 & 0.937 & 0.013 & 0.897 & 0.053 \\ 
		& 15 & 0.945 & 0.005 & 0.908 & 0.042 \\ \hline 
		0.995 & 5 & 0.982 & 0.013 & 0.945 & 0.050 \\ 
		& 10 & 0.989 & 0.006 & 0.952 & 0.043 \\ 
		& 15 & 0.989 & 0.006 & 0.956 & 0.039 \\ \bottomrule 
	\end{tabular}
	\label{tab:2014_on_2015}
\end{table}

\begin{table}
    \scriptsize
    \caption{Results of risk-adjusted incidence model trained from 2012-2014 CFFPR data validated on 2012-2014 (top) and 2015 (bottom) data, where $\alpha$ is the significance level, $m$ is the number of blocks in block jackknife estimation of variance, Coverage is the proportion of programs whose true number of validation incidence cases is contained in the risk-adjusted $1-\alpha$ confidence interval trained from the training data, and AbsCovDiff is the absolute difference between meanCoverage and $1-\alpha$.}
	\centering
	\begin{tabular}{rr||cc|cc}\toprule  
		& & \multicolumn{2}{c|}{PA} & \multicolumn{2}{c}{MRSA} \\ \toprule
		$1-\alpha$ & m & Coverage & AbsCovDiff & Coverage & AbsCovDiff \\ \hline 
		\multicolumn{6}{c}{2012-2014 Training, 2012-2014 Validation} \\ \hline 
		0.7 & 5 & 0.590 & 0.110 & 0.598 & 0.102 \\ 
		& 10 & 0.638 & 0.062 & 0.635 & 0.065 \\ 
		& 15 & 0.668 & 0.032 & 0.683 & 0.017 \\ \hline 
		0.8 & 5 & 0.690 & 0.110 & 0.697 & 0.103 \\ 
		& 10 & 0.727 & 0.073 & 0.731 & 0.069 \\ 
		& 15 & 0.782 & 0.018 & 0.771 & 0.029\\ \hline 
		0.9 & 5 & 0.830 & 0.070 & 0.815 & 0.085 \\ 
		& 10 & 0.863 & 0.037 & 0.845 & 0.055 \\ 
		& 15 & 0.893 & 0.007 & 0.882 & 0.018 \\ \hline 
		0.95 & 5 & 0.900 & 0.050 & 0.878 & 0.072 \\ 
		& 10 & 0.915 & 0.035 & 0.904 & 0.046 \\ 
		& 15 & 0.934 & 0.016 & 0.930 & 0.020 \\ \hline 
		0.995 & 5 & 0.959 & 0.036 & 0.974 & 0.021 \\ 
		& 10 & 0.963 & 0.032 & 0.978 & 0.017 \\ 
		& 15 & 0.974 & 0.021 & 0.982 & 0.013 \\ \hline 
		\multicolumn{6}{c}{2012-2014 Training, 2015 Validation} \\ \hline
		0.7 & 5 & 0.893 & 0.193 & 0.874 & 0.174 \\ 
		& 10 & 0.930 & 0.230 & 0.893 & 0.193 \\ 
		& 15 & 0.937 & 0.237 & 0.904 & 0.204 \\ \hline 
		0.8 & 5 & 0.941 & 0.141 & 0.933 & 0.133 \\ 
		& 10 & 0.948 & 0.148 & 0.937 & 0.137 \\ 
		& 15 & 0.952 & 0.152 & 0.944 & 0.144\\ \hline 
		0.9 & 5 & 0.974 & 0.074 & 0.967 & 0.067 \\ 
		& 10 & 0.974 & 0.074 & 0.967 & 0.067 \\ 
		& 15 & 0.974 & 0.074 & 0.970 & 0.070 \\ \hline 
		0.95 & 5 & 0.981 & 0.031 & 0.978 & 0.028 \\ 
		& 10 & 0.981 & 0.031 & 0.978 & 0.028 \\ 
		& 15 & 0.985 & 0.035 & 0.978 & 0.028 \\ \hline 
		0.995 & 5 & 0.989 & 0.006 & 0.989 & 0.006 \\ 
		& 10 & 0.989 & 0.006 & 0.989 & 0.006 \\ 
		& 15 & 0.989 & 0.006 & 0.993 & 0.002 \\ \bottomrule 
	\end{tabular}
	\label{tab:2012-2014_on_2015_mrsa}
\end{table}

\begin{figure}[htpb!]
	\centering 
	\includegraphics[scale=0.2]{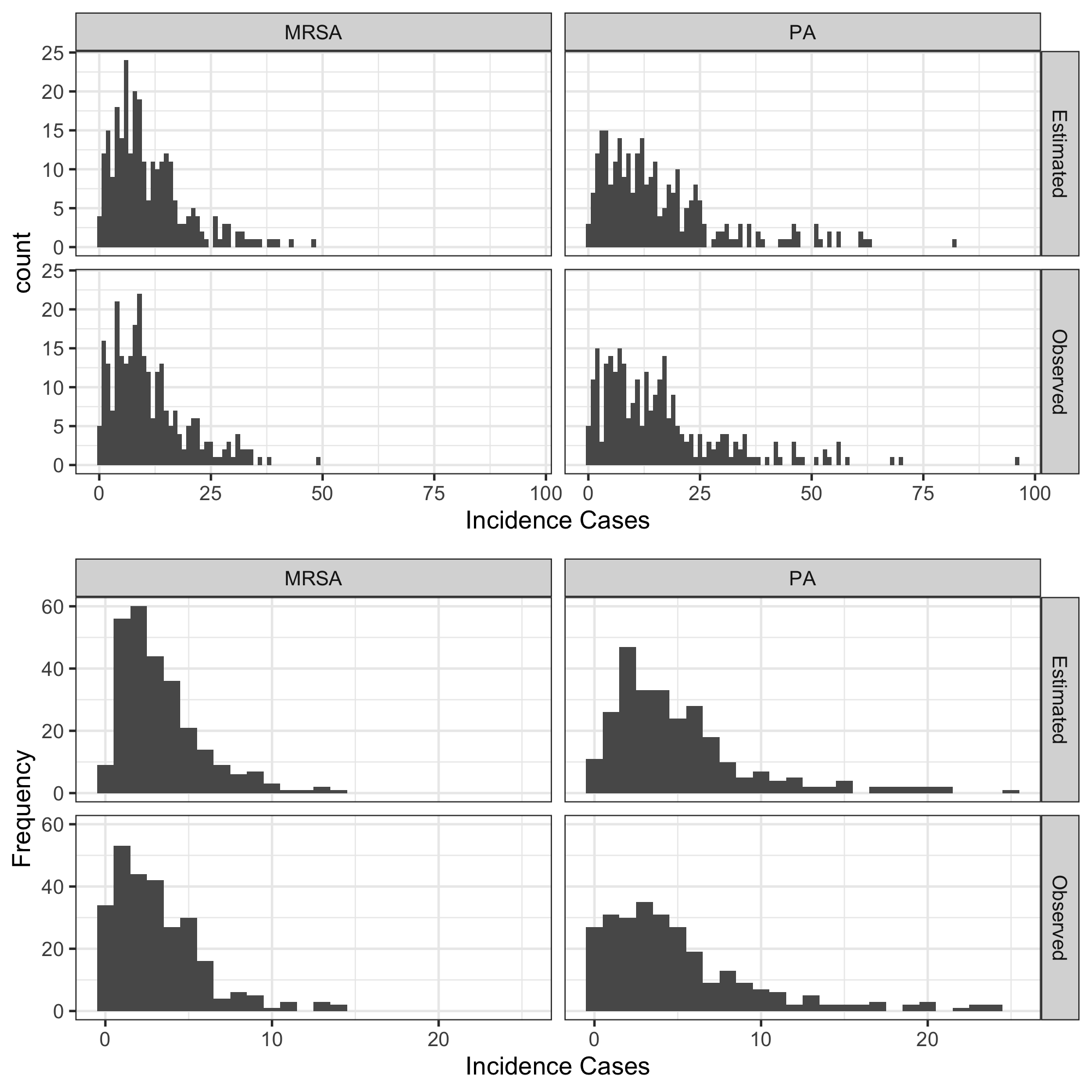}
	\caption{Histogram of estimated versus observed number of 2012-2014 (top) and 2015 (bottom) MRSA and PA incidence cases where the risk-adjusted model is trained with 2012-2014 data}
	\label{fig:2012-2014_on_2015}
\end{figure}

Distribution comparison between estimated and observed survival events for both MRSA and PA is visualized as histograms in Figure 1 \ref{fig:2012-2014_on_2015}. 
The top figure has more incidence cases because it has three years of data (2012-2014) compared with the bottom figure which has only 2015 data. 
We concluded that the estimated and observed distributions were both skewed to the right with similar modes and tails, regardless of bacteria and training year. The estimated incidence distribution for the 2015 validation (bottom figure) has a lower peak compared with the observed distribution for both bacteria due to regression to the mean. Additionally, the estimated Spearman correlation coefficients are $0.87$ (MRSA) and $0.91$ (PA) for 2012-2014 validation and $0.58$ (MRSA) and $0.80$ (PA) for 2015 validation, all of which are moderate to strong correlations, with better validation results in PA. Overall, the correlations are stronger for CF data than simulated data and we speculate this is due to model settings, parameter estimation, and how much variance can be explained by the covariates.

\begin{sidewaysfigure}[htpb!]
    \centering
    \includegraphics[scale=0.5]{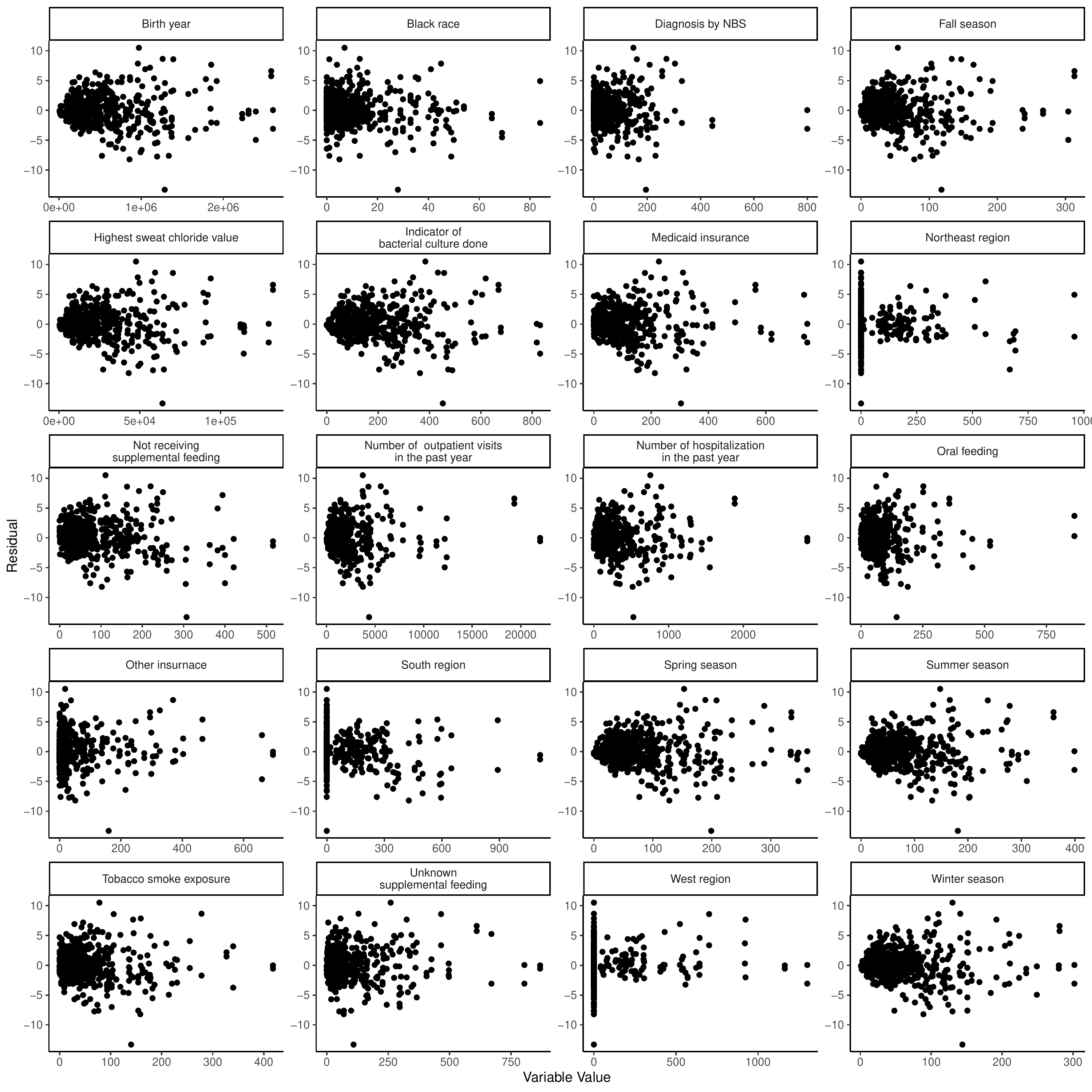}
    \caption{MRSA residual plot of each center-level covariate weighted by at-risk patient time with 2014 data as training set and 2015 data as validation set ($m=15$)}
    \label{fig:residual_mrsa_2014_on_2015}
\end{sidewaysfigure}

\begin{sidewaysfigure}[htpb!]
    \centering
    \includegraphics[scale=0.5]{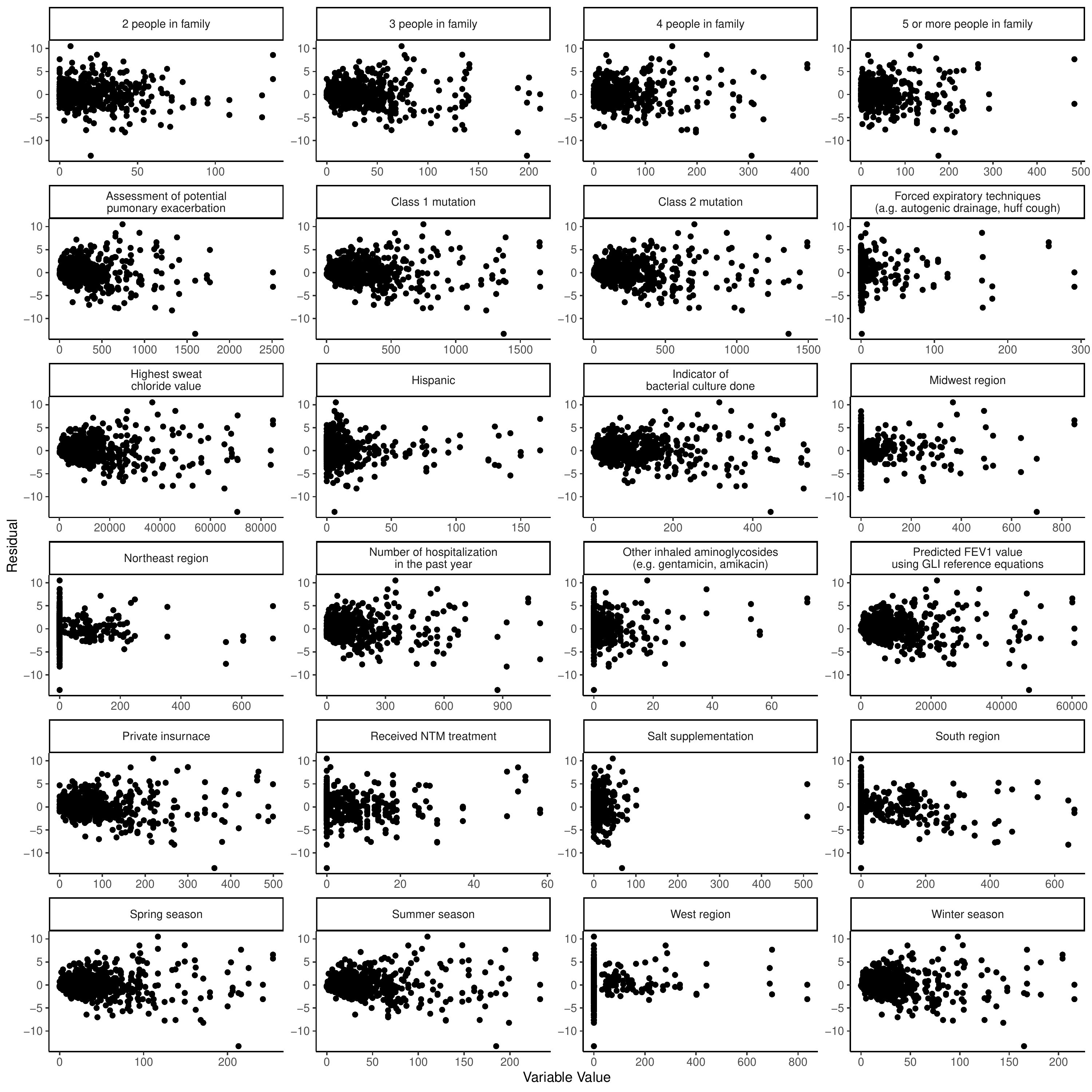}
    \caption{PA residual plot of each center-level covariate weighted by at-risk patient time with 2014 data as training set and 2015 data as validation set ($m=15$)}
    \label{fig:residual_pa_2014_on_2015}
\end{sidewaysfigure}

There are many other ways to investigate model adequacy. An alternative way to evaluate goodness-of-fit of our risk adjusted model is residual plot. After the model was trained with the training set, residuals of the number of incidence events can be calculated in the validation set. Each center-level covariate was summarized across patients and weighted by their length of at-risk time. The results of 2014 training data and 2015 validation data are shown in Figures \ref{fig:residual_mrsa_2014_on_2015} and \ref{fig:residual_pa_2014_on_2015}
for MRSA and PA respectively. No obvious patterns were detected across values of each weighted covariate and most residuals are evenly distributed across 0. This implies that there is no clear evidence of poor model fit.

\clearpage
\newpage

\section*{Appendix D: Glossary for Key Notations Introduced in the Methods}\label{supp3:glossary}

\begin{enumerate}

\item $b$: The indicator of a block in the block jackknife. 

\item $\beta_0$: The covariate coefficient in the intensity process $R_{0j}$.


\item $\hat{\bar\beta}_{M}$: The pooled estimator of $\beta_0$ averaged across all $M$ $\hat{\beta}_{nl}$'s where $l = 1, \ldots, M$.

\item $\hat{\beta}_{nl}$: The estimator of $\beta_0$ from the mixed effect AG model for the $l$th multiple imputation dataset of size $n$. 

\item ${\hat{\beta}_{nl}}^{(-v)}$: The estimator of $\beta_0$ from the mixed effect AG model for the $l$th multiple imputation dataset of size $n$ but with the $v$th level-1 group taken out.

\item $c_j$: Total number of unique level-2 group variable (i.e., $i$) in the $j$th level-1 group.

\item $D$: The original data.

\item $D^*$: The block jackknife data, derived from the original data. 

\item $D^{bs}$: The bootstrapped data with the replacement of level-1 groups, derived from the original data. 

\item $d\hat{\Lambda}_{nl}(t)$: The estimator of the hazard function at time $t$ for the $l$th multiple imputed data.


\item $d\hat\Lambda_{nl}^{(-v)}(t)$: The estimator of the hazard function at time $t$ for the $l$th multiple imputation data but with the $v$th level-1 group taken out. 

\item $dN_{ijk}(t)$: The indicator of an observed survival event at time $t$ for the $j$th level-1 group, $i$th level-2 group, and $k$th encounter.

\item $i$: The index for level-2 group variables in the survival data with a two-level hierarchical structure, e.g., CF patient (the second level).

\item $j$: The index for level-1 group variables in the survival data with a two-level hierarchical structure, e.g., CF program (the first level).

\item $k$: The index for observations in a level-2 group, e.g., each encounter at a CF program.

\item $L_{ijk}$: The lower bound of the at-risk interval for the $j$th level-1 group, $i$th level-2 group, and $k$th encounter; used in $d\hat{\lambda}_{nl}(s)$. 

\item $l$: The index for multiple imputation datasets.

\item $M$: Total number of multiple imputation datasets. 

\item $m$: Total number of blocks in the block jackknife. 

\item $m_{ij}$: Total number of observations in the $i$th level-2 group and the $j$th level-1 group.

\item $N$: Total number of unique level-1 group variables (indexed by $j$).

\item $N_j$: The observed number of survival events for the $j$ level-1 group.

\item $\hat{N}_{jl}$: The estimated number of survival events for the $j$th level-1 group and $l$th multiple imputed dataset. 

\item $\hat{N}_{j\bigcdot}$: The estimated number of survival events for the $j$th level-1 group across $M$ multiple imputation datasets.

\item $\hat{N}_j \l({\hat{\beta^*}}^{(-b)}, {d\hat{\Lambda^*}}^{(-b)} \r)$: The number of estimated survival events for the $j$th level-1 group based on estimated parameters of data with the $b$th block removed.

\item $\hat{N}_{jl}^{(-v)}$: The estimated number of survival events for the $j$th level-1 group based on estimated parameters for the $l$ multiple imputation dataset with the $v$th level-1 group removed.

\item $n$: Total sample size, $n = \sum_{j=1}^N \sum_{i=1}^{c_j} m_{ij} $. 

\item $p$: The dimension of $\Z_{ijk}(t)$; the number of all covariates of interest. 

\item $q_{m,N}$: The number of elements in each block of the block jackknife method for $m$ blocks and sample size $N$. 

\item $R_{0j}$: The true intensity process for the $j$th level-1 group. 

\item $S^*_{j}$: The estimated variance of the survival events for the $j$th level-1 group pooled across all $m$ block jackknife datasets.

\item $S^*_{jl}$: The estimated variance of the survival events for the $j$th level-1 group pooled across all $m$ block jackknife datasets using the $l$th multiple imputation data as the original data.

\item $\hat{s}^2_{j\bigcdot}$: The estimated multiple imputation variance in the $j$th level-1 group.

\item $T_{ijk}$: The at-risk time interval for the $k$th encounter, the $i$th level-2 group, and the $j$th level-1 group.

\item $\hat{T}_j$: The test statistic of the Z-test comparing the estimated number of events with the observed number of events for the $j$th level-1 group. 

\item $t$: Time. 

\item ${{\hat{\theta^*}}_j}^{(-b)}$: The estimator of the survival events for the $j$th level-1 group based on estimated parameters of data with the $b$th block removed.

\item $\bar{\theta}^*_j$: The estimated survival events pooled across all $m$ block jackknife datasets.

\item $U_{ijk}$: The upper bound of the at-risk interval for the $j$th level-1, $i$th level-2, and $k$th encounter; used in $d\hat{\lambda}_{nl}(s)$. 

\item $\hat{\widetilde{V}}_{j}$: The estimated variance of the estimated number of survival events $\hat{N}_{j \bigcdot}$. 

\item $\hat{V}_{j\bigcdot}$: The estimated across-group variance in the $j$th level-1 group.

\item $Var\left(\hat{\bar\beta}_{M}\right)$: The pooled estimator of the variance of $\beta_0$ across $M$ multiple imputed datasets based on $\hat{\beta}_{nl}$. 

\item $v$: Indicator of the level-1 group that is taken out of training set in the jackknife method.

\item $\Z_{ijk}(t)$: The $p$-dimensional covariate variable at time $t$ for the $k$th encounter, the $i$th level-2 group, and the $j$th level-1 group, including both time-invariant and time-varying variables.

\item $\Z_{ijkl}(t)$: The $p$-dimensional covariate variable in the $l$th multiple imputed dataset at time $t$ for the $k$th encounter, the $i$th level-2 group, the $j$th level-1 group, including both time-invariant and time-varying variables.
\end{enumerate}

\clearpage
\bibliographystyle{apalike}
\bibliography{survivalbib}
\end{document}